\documentclass[Afour,sagev,times]{sagej}

\usepackage{moreverb,url}

\usepackage{amsmath,amsfonts}
\usepackage{algorithmic}
\usepackage{array}
\usepackage{kotex}
\usepackage{xspace}
\usepackage{colortbl}
\usepackage[table]{xcolor}
\usepackage{verbatim}
\usepackage{graphicx}

\usepackage{multicol}
\usepackage{multirow}
\usepackage{listings}

\usepackage[colorlinks,bookmarksopen,bookmarksnumbered,citecolor=red,urlcolor=red]{hyperref}

\newcommand\BibTeX{{\rmfamily B\kern-.05em \textsc{i\kern-.025em b}\kern-.08em
T\kern-.1667em\lower.7ex\hbox{E}\kern-.125emX}}

\def\volumeyear{2016}

\begin{document}

\runninghead{Jung et al.}

\title{Seeing Graphs Like Humans: Benchmarking Computational Measures and MLLMs for Similarity Assessment}

\author{Seokweon Jung\affilnum{1}\affilnum{2}, Jeongmin Rhee\affilnum{3}, Seoyoung Doh\affilnum{4}, Hyeon Jeon\affilnum{2}, Ghulam Jilani Quadri\affilnum{5} and Jinwook Seo\affilnum{2}}

\affiliation{
\affilnum{1}LLM Innovation Research Center, Korea Advanced Institute of Science and Technology (KAIST), Daejeon,  Republic of Korea\\
\affilnum{2}Department of Computer Science and Engineering, Seoul National University, Seoul, Republic of Korea\\
\affilnum{3}Interdisciplinary Program in Artificial Intelligence, Seoul National University, Seoul, Republic of Korea\\
\affilnum{4}College of Liberal Studies, Seoul National University, Seoul, Republic of Korea\\
\affilnum{5}School of Computer Science, University of Oklahoma, Norman, Oklahoma, United States
}

\corrauth{Jinwook Seo, HCI Lab, Department of Computer Science and Engineering, Seoul National University, 1 Gwanak-ro, Gwanak-gu, Seoul 08826, Republic of Korea}
\email{jseo@snu.ac.kr}

\newcommand{\ovs}{\textit{Overall Shape}\xspace}
\newcommand{\lcs}{\textit{Local Shapes}\xspace}
\newcommand{\gsz}{\textit{Graph Size}\xspace}
\newcommand{\eds}{\textit{Edge Density}\xspace}
\newcommand{\ndg}{\textit{Node Degrees}\xspace}
\newcommand{\cmm}{\textit{Communities}\xspace}

\newcommand{\revised}[1]{\textcolor{red}{#1}}

\begin{abstract}
Comparing graphs to identify similarities is a fundamental task in visual analytics of graph data. To support this, visual analytics systems frequently employ quantitative computational measures to provide automated guidance. However, it remains unclear how well these measures align with subjective human visual perception, thereby offering recommendations that conflict with analysts' intuitive judgments, potentially leading to confusion rather than reducing cognitive load. Multimodal Large Language Models (MLLMs), capable of visually interpreting graphs and explaining their reasoning in natural language, have emerged as a potential alternative to address this challenge.
This paper bridges the gap between human and machine assessment of graph similarity through three interconnected experiments using a dataset of 1,881 node-link diagrams. Experiment 1 collects relative similarity judgments and rationales from 32 human participants, revealing consensus on graph similarity while prioritizing global shapes and edge densities over exact topological details. Experiment 2 benchmarks 16 computational measures against these human judgments, identifying Portrait divergence as the best-performing metric, though with only moderate alignment. Experiment 3 evaluates the potential of three state-of-the-art MLLMs (GPT-5, Gemini 2.5 Pro, Claude Sonnet 4.5) as perceptual proxies. The results demonstrate that MLLMs, particularly GPT-5, significantly outperform traditional measures in aligning with human graph similarity perception and provide interpretable rationales for their decisions, whereas Claude Sonnet 4.5 shows the best computational efficiency. Our findings suggest that MLLMs hold significant promise not only as effective, explainable proxies for human perception but also as intelligent guides that can uncover subtle nuances that might be overlooked by human analysts in visual analytics systems.
\end{abstract}


\maketitle


\section{Introduction}
Comparing multiple graphs visually to discern structural variations and topological shifts is a fundamental task in visual analytics across diverse domains~\cite{comparison-biology, comparison-transportation, sns-measurecomparison}, enabling analysts to discern dominant trends or outlying patterns. This is particularly critical in dynamic graph analysis~\cite{temporalnetwork}, where understanding temporal evolution commonly involves slicing dynamic graphs into multiple static snapshots and assessing their similarities and differences over time~\cite{dynamicnetwork-visualization, beckTaxonomySurvey2017}. To support this, modern visual analytics systems typically employ a dual approach: leveraging human visual perception for qualitative interpretation while utilizing quantitative computational measures to reduce cognitive load and guide the user's attention~\cite{dgcomics, diffseer, monetexplorer, jeon24tvcg}.

While this synergistic approach aims to mitigate the complexity of analysis, its foundational premises remain insufficiently validated in graph comparison. In other domains, such as scatterplot analysis, the alignment between human perception and computational metrics has been extensively studied to optimize system design~\cite{tatuVisualQuality2010, albuquerquePerceptionbasedVisual2011}. In contrast, although previous studies have proposed design guidelines for comparison tasks~\cite{perceptualproxies, considerations-comparison}, the extent to which humans can effectively perceive differences between graphs remains underexplored~\cite{ghulam-survey}. 
Consequently, to design effective visual analytics systems, it is imperative to first establish a baseline understanding of the environments and support mechanisms that facilitate human graph comparison.

Graph comparison is compounded by diverse properties such as size, density, and layout~\cite{survey-emprical-graphvis, bartolomeo-graph-layout-evalutaion}, alongside critical factors like visual scalability and mental map preservation~\cite{hermanGraphVisualization2000, mentalmap-preservation}.
This complexity makes it unclear which graph features or visual cues humans prioritize when assessing similarity, and whether these perceptual judgments are consistent across different observers.
Therefore, we propose our first research question: \textit{Can humans discern similarity and differences between graph visualizations, and which factors significantly affect these decisions?}

To complement human judgment, computational measures offer objective quantification. While a wide range of metrics exists, from classical graph edit distances to advanced methods such as spectral distances~\cite{gao2010survey, weller2015survey, soundarajan2014survey}, their alignment with human visual perception remains largely unverified.
This lack of empirical validation poses a critical risk: automated systems may offer recommendations that conflict with analysts' intuitive judgments, potentially leading to confusion and mistrust rather than reducing cognitive load~\cite{jeon25arxiv, jeon25chi, shneiderman22book}.
This knowledge gap leads to our second research question: \textit{Do traditional computational measures capture the similarities and differences that humans perceive visually?}

Seeking a computational method that better aligns with human intuition, we examine methodologies capable of interpreting the holistic and semantic context of graph visualizations.
Recent advances in Multimodal Large Language Models (MLLMs) have emerged as a computational method for supporting visual analytics tasks by demonstrating notable capabilities in interpreting visual content, including charts and diagrams~\cite{gpt-visdesign, viseval, bavisitter}.
Although some research suggests that the MLLMs are nearly blind at comprehending complex visualizations~\cite{vlm-blind}, recent studies indicate they are overcoming these barriers, demonstrating promising capabilities in layout generation and basic perception~\cite{gpt-graphdrawing, fanHowWell2025, millerExploringMLLMs2025}.
Despite these advancements, their utility in graph comparison remains underexplored, yet their evolving visual understanding suggests they hold significant potential to act as proxies for human perception~\cite{gpt-vis-future}.
Based on this potential, we propose our third research question: \textit{Do MLLMs possess the capability to align with human perception and guide users in graph comparison tasks?}

To address these three research questions, we conduct a series of three interconnected experiments. 
To validate our questions under diverse conditions, we create a comprehensive dataset comprising 1,881 node-link diagrams. These are generated from both synthetic and real-world data, systematically varying in size, density, and layout, which are graph attributes known to have great influence on human perception of graph visualizations~\cite{huangMeasuringEffectiveness2009, effect-of-graphlayout}.

In Experiment 1, to account for the subjective nature of human similarity perception, we employ a relative-comparison task in which participants judge which of two target graphs is more similar to a reference graph~\cite{clusetme, comparing-similarity-timeseries}. We collect 2,208 relative similarity judgments, confidence ratings, and corresponding explanations from 32 participants. Our analysis reveals consistent trends in human perception of graph similarity across varying graph conditions, providing insights into the factors influencing both perception and self-assessed certainty.

In Experiment 2, we evaluate the extent to which computational graph similarity measures align with human perceptual judgments. We select 16 established measures representing a range of theoretical backgrounds based on a comprehensive literature review~\cite{soundarajan2014survey, emmert2016survey, donnat2018survey, masuda2019survey, wills2020survey, tantardini2019survey}. We assess agreement between human choices and computational scores, identify which measures most closely reflect human judgments, and analyze how discrepancies relate to participant confidence.

Finally, in Experiment 3, we explore the potential of three state-of-the-art MLLMs, GPT-5, Gemini 2.5 Pro, and Claude Sonnet 4.5, to visually assess graph similarity. The models are prompted to perform the same relative comparison task as human participants, including providing confidence ratings and explanations for their decisions. Our analysis demonstrates that MLLMs exhibit a higher overall agreement with human judgments than traditional computational measures. Furthermore, they provide interpretable explanations for their decision criteria, suggesting they can effectively lower the barrier to visual analytics of graph data.

Synthesizing the insights from this series of experiments, we propose evidence-based guidelines for designing visual analytics systems that include graph comparison tasks, such as observing temporal evolution or detecting anomalies in dynamic graphs. We outline how to effectively structure graph comparison tasks for human analysts, select perception-aligned computational metrics, and strategically leverage MLLMs to provide explainable guidance, ultimately aiming to bridge the gap between human intuition and machine quantification.

In summary, this work makes the following contributions:
\begin{itemize}
    \item We construct a large-scale dataset of graph visualizations systematically varied by topological and visual properties to benchmark similarity perception.
    \item We provide empirical evidence of consistent patterns in human perception of graph similarity and identify key factors influencing these judgments.
    \item We identify the specific computational similarity measures that best approximate human visual perception of graph differences.
    \item We demonstrate that state-of-the-art MLLMs closely align with human perception and provide interpretable rationales, highlighting their potential as effective assistants in visual analytics.
\end{itemize}

\section{Related Work}
We discuss relevant literature on computational measures for graph comparison as well as visualization techniques for comparative analysis tasks involving graphs.

\subsection{Computational Graph Comparison}
\subsubsection{Measure-based Comparison}
~\label{sec:rw-measures}
Graphs can be quantitatively compared using graph similarity measures, often referred to as graph distances. Various graph distance measures have been developed to facilitate graph comparisons. In our study, we review literature published in SCI(E) journals and presented at leading international conferences over the past decade. Our investigation included surveys on graph and network similarity/distance that compared multiple measures~\cite{soundarajan2014survey, emmert2016survey, donnat2018survey}, as well as studies that evaluated and contrasted various similarity measures to provide further research guidance~\cite{masuda2019survey, wills2020survey, tantardini2019survey}. This literature review enables us to both classify the existing graph similarity measures and identify those applicable to our target graphs.

A key classification criterion, as highlighted in the survey studies, is node correspondence~\cite{soundarajan2014survey, emmert2016survey, donnat2018survey}. When graphs contain node labels and a one-to-one correspondence exists between nodes, this information is critical for node-based comparisons. Measures such as graph edit distance~\cite{gao2010survey} and DELTACON~\cite{deltacon} are designed for scenarios with known node correspondence (KNC). However, as noted in our \hyperref[sec:studydesign]{study design}, our target graphs are characterized by unknown node correspondence (UNC), thus we exclude KNC-based methods from our analysis.

UNC graph similarity measures determine similarity based on high-level structural features rather than relying on node-specific details. Depending on the granularity of the features considered, these measures are typically categorized with the levels of granularity: local and global~\cite{soundarajan2014survey, emmert2016survey, donnat2018survey}. 

The applicability and performance of these measures vary significantly depending on the specific characteristics of the graphs, such as whether they are directed, weighted, or contain self-loops~\cite{tantardini2019survey}. Moreover, frameworks have categorized these similarity measures based on their scope, distinguishing them as micro-level, meso-level, and macro-level measures~\cite{donnat2018survey}. Several studies extensively reviewed and compared these diverse graph similarity metrics to better understand their advantages, limitations, and practical applicability~\cite{tantardini2019survey, wills2020survey, masuda2019survey}. 

However, despite extensive comparative analyses of computational measures, the literature currently lacks empirical studies investigating how these measures align with human perceptions when the results are visually presented.
To address this gap, we select representative computational measures suited to our experimental context.
These measures span diverse theoretical backgrounds~\cite{emmert2016survey}, ranging from straightforward approaches such as Jaccard distance and edit distance~\cite{gao2010survey} to more sophisticated spectral similarity measures~\cite{graphspectra}, graph kernels~\cite{graphkernels}, and graphlet-based techniques~\cite{gcd}.

\subsubsection{Human Visual Comparison}
Another important method for comparing graphs is visual analysis, which helps users directly identify similarities and differences through graphical representations~\cite{visualcomparison-infovis}. Effective visual comparison requires careful consideration of visualization principles and perceptual guidelines~\cite{perceptualproxies, considerations-comparison}.

Visual comparison methods are particularly prevalent in analyzing dynamic graphs, which frequently represent complex and large-scale temporal datasets~\cite{dynamicnetwork-visualization, temporalnetwork}. Taxonomies of dynamic graph inspection explicitly include comparative analysis tasks~\cite{tasktaxonomy-temporalgraph}. Graph visualization techniques are extensively employed to present both overall patterns~\cite{dgcomics, diffseer, monetexplorer} and intricate details~\cite{adamotif, stagedanimation, diffseer, monetexplorer} of graph data.

The comparison itself is recognized as a foundational, low-level visual analytic activity~\cite{low-level-va-activity} involving the examination of attributes and relationships within data. Previous studies have investigated the perceptual aspects of graph visualizations, typically focusing on the readability or performance of different visualization techniques~\cite{ghulam-survey, ghoniem2004comparison, guo2015representing, chang2017evaluating}. More recently, research has examined how different node-link layouts influence the perception of graph properties~\cite{effect-of-graphlayout}. However, studies explicitly investigating human perception of graph similarity remain limited.

Empirical studies using graph data have highlighted substantial variability in perceptual performance and preferences according to graph size and complexity~\cite{soundarajan2014survey, burch-graph-empirical-evaluation}. 
One notable study incrementally adjusted simple graph attributes to evaluate perceptual thresholds for detecting graph differences~\cite{investigatin-gs-perception}. Although there have been empirical examinations of perceptual differences related to specific graph features~\cite{effect-of-graphlayout, mooney-graphdrawing}, none have explicitly targeted similarity perception. Research examining temporal network changes has primarily focused on mental map preservation rather than similarity~\cite{mentalmap-preservation}. 
In our study, we quantitatively investigate the relationship between human visual perception and computational graph similarity measures.

\subsection{Quantifying Visual Perception}
Quantifying human perception, which is an inherently subjective phenomenon, has been explored across various visualization contexts~\cite{perceptualproxies, considerations-comparison}. Researchers have developed methods to measure perceptual judgments in different visualization domains, including visual encoding~\cite{sepme, channel-effectivness}, clustering~\cite{ghulam-clusterperception, clusetme, revisiting-dr}, and time-series data~\cite{comparing-similarity-timeseries}.

While perceptual tasks involving node-link diagrams and adjacency matrices have also been studied~\cite{mapping-perceptual-network}, specific research on human perception of graph similarity is notably scarce. Existing related work primarily focused on the perceptual effects of varying graph properties or visualization techniques rather than explicitly addressing similarity perception~\cite{effect-of-graphlayout}. Our research aims to fill this gap by explicitly measuring perceptual similarity judgments in graph visualizations.

\begin{figure*}[t]
  \centering
  \includegraphics[width=\textwidth, alt={methodsummary}]{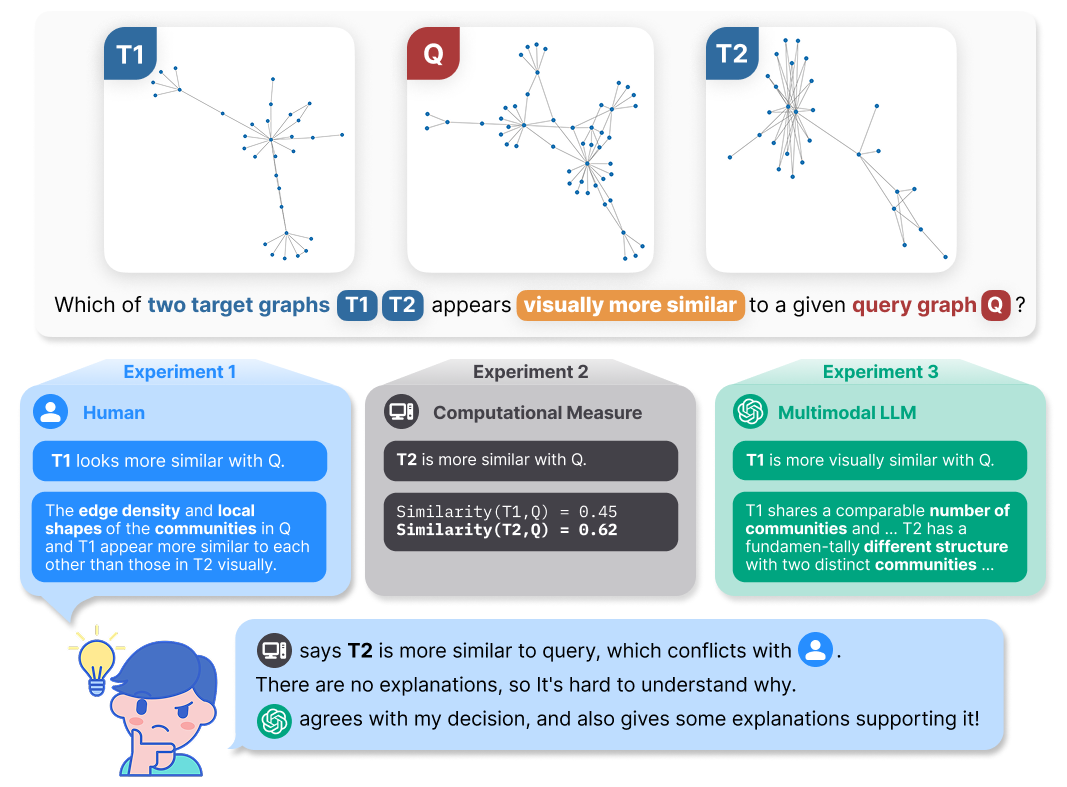}
  \caption{An overview of our research methodology, structured into three interconnected experiments designed to investigate graph comparison capabilities in humans and machines. Experiment 1 assesses human competence in graph similarity assessment through indirect similarity measurements derived from visual perception (RQ1). Experiment 2 computes pairwise similarities using 16 distinct computational graph similarity measures and compares them with human decisions to determine the alignment between humans and machines (RQ2). Experiment 3 evaluates the relative graph similarity assessment capabilities of MLLMs, analyzing both their alignment with human perception and the interpretability of their reasoning (RQ3). Our results demonstrate that MLLMs exhibit a higher alignment with human judgment than computational measures, qualifying them as superior perceptual proxies. Furthermore, by providing interpretable decision rationales, they serve as a more effective method for assisting human analysts in graph comparison tasks.
  }
  \label{fig:method}
\end{figure*}

\subsection{MLLMs for Graph Visualization}
The emergence of powerful Multimodal Large Language Models (LLMs) has enabled the interpretation and generation of visual content through natural language descriptions~\cite{llm-ready-vis, gai-for-visualization, shpark-genui, shpark-search}.
While some studies have characterized MLLMs as being nearly ``blind'' in their visualization interpretation capabilities due to early limitations~\cite{vlm-blind}, their potential remains substantial, and further advancements are expected to significantly impact future visualization research~\cite{gpt-vis-future}. 
A growing body of work demonstrates that state-of-the-art models are overcoming these initial barriers, exhibiting significant potential in both generative and perceptual tasks. 
For instance, they have been utilized to create visualization guidelines~\cite{bavisitter, gpt4-mllm-visualization-literacy, choeEnhancingData2025}, evaluate visualization effectiveness~\cite{viseval}, and generate novel visualizations~\cite{gpt-visdesign}.

Specifically in the domain of graph visualization, although the inherent complexity of graph topology poses a unique barrier, recent improvements in model performance have spurred research into leveraging MLLMs for diverse graph-related tasks.
Di Bartolomeo et al. explored the generative capabilities of ChatGPT for graph drawing, revealing that while challenges remain in handling complex constraints, the model exhibits significant potential in applying layout algorithms and coordinating spatial arrangements via text prompts~\cite{gpt-graphdrawing}.
Building on this, Fan et al. conducted a comprehensive evaluation of LLMs on graph layout tasks, reporting that advanced models, such as GPT-4, can effectively adhere to aesthetic constraints and optimize layout metrics, particularly for small-scale graphs~\cite{fanHowWell2025}. 
Shifting the focus from generation to perception, Miller et al. investigated MLLMs' understanding of network visualization principles~\cite{millerExploringMLLMs2025}. Their findings suggest that these models are beginning to align with human judgments regarding fundamental design aspects, such as visual clutter and Gestalt principles.

Despite these significant advancements in layout generation and single-graph assessment, the specific task of graph comparison remains underexplored in the existing literature. 
Comparing graphs is a foundational analytic activity, essential not only for distinguishing structural differences between static graphs but also for analyzing temporal evolution and detecting anomalies in dynamic graph analysis.
In our study, we explore the extent to which current MLLMs comprehend graph similarity, investigate their decision criteria, and evaluate the consistency of their assessments relative to human judgments and computational measures.

\section{Methodology}
\label{sec:methodology}
We conduct a series of three interconnected experiments to provide empirical insights for designing effective visual analytics systems in which users must compare graphs across multiple graph visualizations. Our study investigates three primary research questions:
\begin{itemize}
    \item \textbf{RQ1:} Can humans discern similarity and differences between graph visualizations, and which factors significantly affect these decisions?
    \item \textbf{RQ2:} Do traditional computational measures capture the similarities and differences that humans perceive visually?
    \item \textbf{RQ3:} Do MLLMs possess the capability to align with human perception and guide users in graph comparison tasks?
\end{itemize}

To address these questions systematically, we first construct a comprehensive dataset of graph visualizations. 
Subsequently, we systematically control graph size, edge density, and layout algorithms, employing them as independent variables known to influence empirical perception~\cite{survey-emprical-graphvis, mooney-graphdrawing, effect-of-graphlayout}.

\hyperref[sec:ex1]{Experiment 1} then uses this dataset to assess how human participants perceive similarities between graphs~\textbf{(RQ1)}. To mitigate individual differences in perceptual sensitivity and the inherent difficulty of quantifying similarity on an absolute scale, we employ a relative comparison task~\cite{comparing-similarity-timeseries}. We develop a web-based experimental interface and collect a total of 2,208 responses from 32 participants.
The subsequent \hyperref[sec:ex2]{Experiment 2} and \hyperref[sec:ex3]{Experiment 3} focus on identifying which computational methods best align with the human similarity assessments collected in Experiment 1~\textbf{(RQ2, RQ3)}. We view this alignment process as the evaluation of the accuracy of computational measures that proxy human perception, adopting a data-driven approach to evaluate Visual Quality Measures (VQMs) based on perceptual data~\cite{datadriven-evaluation}.

\begin{table}[t]
    \small\sf\centering
    \caption{Three independent variables. We utilize the notion of linear edge density (\(\frac{|E|}{|V|}\)) to follow the classification criteria by Yoghourdjian et al.~\cite{survey-emprical-graphvis}. Layout algorithms are selected based on popularity and ease of use~\cite{effect-of-graphlayout}.}
    \begin{tabular}{ m{0.8cm} >{\raggedright\arraybackslash}m{1.7cm} >{\raggedright\arraybackslash}m{4.4cm} }
    \toprule
     \textbf{Var.} & \textbf{Description} & \textbf{Value range} \\
     \midrule
        Size \newline (\textbf{\textit{N}}) & \footnotesize Number of nodes &  \footnotesize  small:[10, 20], medium:[21, 50], large:[51, 200], very large:[201, 400] \\
     \midrule
        Density \newline (\textbf{\textit{D}}) & \footnotesize Number of edges divided by nodes &  \footnotesize sparse:[1, 2), dense:[2, 3), \newline very dense:[3, 10] \\
     \midrule
        Layout \newline (\textbf{\textit{L}}) & \footnotesize Algorithm to place nodes and edges &    \footnotesize force-directed (F-R), circular, multidimensional scaling (UMAP) \\ 
     \bottomrule
    \end{tabular}
    \label{table:parameters}
\end{table}

\subsection{Study Design} 
\label{sec:studydesign}

\subsubsection{Target Graph Specification}
\label{graphconfig}
To establish robust foundations and provide reliable ground-truth data, our research systematically examines perceived graph similarity using the most fundamental form of graph visualization. We focus on undirected, unweighted graphs without self-loops. Despite their simplicity, these graphs are standard in graph visualization research, and many visualization techniques~\cite{adamotif, timelighting} and similarity measures~\cite{graphletcorrelationdistance, netlsd} are explicitly designed for such configurations. 

Another critical consideration in graph comparison tasks is node correspondence, which can be known node correspondence (KNC) or unknown node correspondence (UNC)~\cite{emmert2016survey}. When node correspondence is known, visualizations typically include node-specific information through additional visual channels, such as color, to represent node attributes. Since our goal is to study basic graph visualizations without the influence of additional node-specific information, we focus on graphs with UNC, thereby avoiding visual cues that might bias assessments.

Moreover, evaluating disconnected graphs requires comparing multiple graphs simultaneously, which raises the difficulty of the comparison task. To maintain consistency and simplicity, we limit our investigation to graphs consisting of a single connected component.


\begin{figure*}[t]
  \centering
  \includegraphics[width=\textwidth, alt={algorithms}]{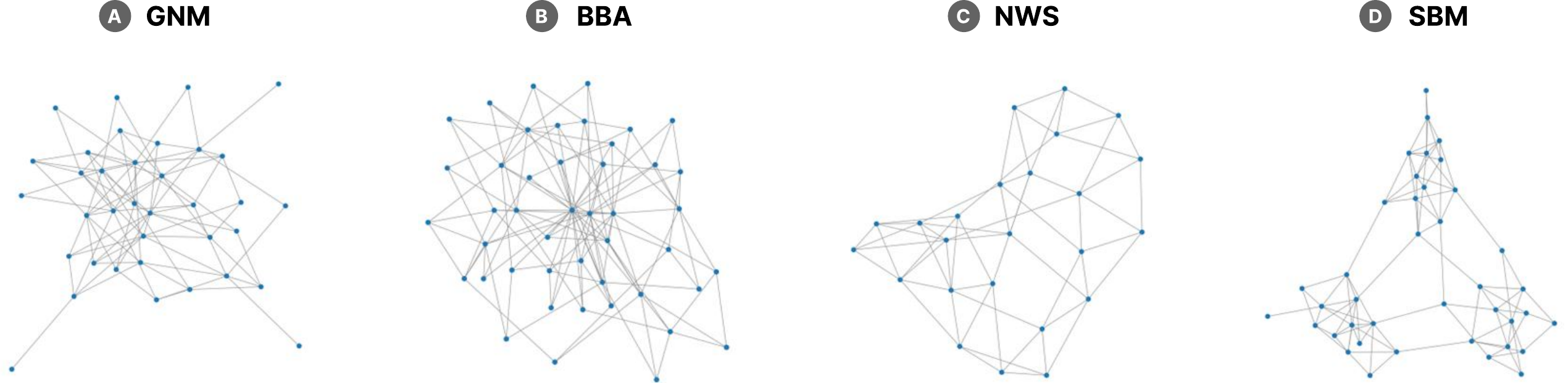}
  \caption{
    In this study, four different synthetic graph generation algorithms are utilized for stimuli generation. A) \textbf{GNM} algorithm randomly generates M edges between N nodes. B) \textbf{BBA} algorithm has power-law degree distributions of node degrees by connecting edges between new nodes to existing nodes with high degree ~\cite{barabasi1999emergence}. C) \textbf{NWS} algorithm creates a ring over nodes and connections between their $k$ nearest neighbors ~\cite{newman1999renormalization}. D) \textbf{SBM} algorithm partitions a graph into blocks of arbitrary sizes whose edges are placed between pairs of nodes ~\cite{holland1983stochastic}.
  }
  \label{fig:synthetic}
\end{figure*}

\subsubsection{Stimuli Generation}
We generate 1,881 node-link diagrams comprising 1,152 synthetic graphs and 729 real-world graphs. Guided by a survey of empirical studies on graph visualization~\cite{survey-emprical-graphvis} and research investigating the factors influencing node-link diagrams~\cite{bartolomeo-graph-layout-evalutaion, mooney-graphdrawing}, the dataset covers a broad spectrum of graph sizes, densities, and layouts. The following subsections detail the generation process.

\paragraph{Data Sources}
Synthetic graphs offer diverse examples, while real-world graphs provide practical relevance~\cite{investigatin-gs-perception}. Thus, we include both data sources for our experiment.
Synthetic graphs are generated using four different graph generation algorithms, all of which generate graphs with distinct characteristics~(\autoref{fig:synthetic}).
\begin{itemize}
    \item \textbf{GNM}: Graphs with randomly generated M edges between N nodes. While the related study~\cite{mooney-graphdrawing} employs the Erdős–Rényi model~\cite{erdHos1966existence}, we utilized GNM to directly control the edge density.
    \item \textbf{BBA (Barabási–Albert)}: Graphs with power-law degree distributions by connecting edges between new nodes to existing nodes with high degree ~\cite{barabasi1999emergence}.
    \item \textbf{NWS (Newman–Watts–Strogatz}: Graphs with a ring over nodes and connection between their $k$ nearest neighbors ~\cite{newman1999renormalization}.
    \item \textbf{SBM (Stochastic Block Model)}: Graphs partitioned into blocks of arbitrary sizes and edges are placed between pairs of nodes ~\cite{holland1983stochastic}.
\end{itemize}

Real-world graphs are extracted from dynamic graph datasets included in SNAP~\cite{leskovec2016snap} and other graph visualization research~\cite{multidynnos}.
We segment each dynamic graph dataset into consecutive static snapshots by slicing it with diverse fixed temporal scales (daily, weekly, monthly, yearly). Datasets that fail the empirical size and density criteria (\autoref{table:parameters}) are excluded, yielding 12 suitable datasets.

\subsubsection{Graph Size (Size)}
Graph size, defined as the number of nodes in a graph, significantly affects visual perception and readability~\cite{ghulam-survey}. To ensure our experiment reflects graph comparison tasks encountered in empirical settings, we adopt the categorization established in a previous survey~\cite{survey-emprical-graphvis}. Accordingly, graph size is categorized into four groups: small ($[10, 20]$), medium ($[21, 50]$), large($[51, 200]$), and very large ($[201, 400]$).
Although the referenced survey~\cite{survey-emprical-graphvis} does not specify an upper bound for the very large category, we impose a limit of 400 nodes to prevent excessive variance in visual complexity within the group.
To mitigate the influence of outliers, graphs within each category are sampled using Gaussian distributions centered on the category median.

\subsubsection{Edge Density (Density)}
Edge density, representing the frequency of connections between nodes, also significantly influences visual perception and readability~\cite{ghulam-survey}. We adhere to the notion of linear density ($|E|/|V|$) and categorization of Yoghourdjian et al.~\cite{survey-emprical-graphvis}, excluding tree-like structures that often result in disconnected components (density range $(0, 1)$). Consequently, the density categories are defined as sparse ($[1, 2)$), dense ($[2, 3)$), and very dense ($[3, 10]$). The upper density limit is capped at 10, approximating the maximum density observed in our real-world dataset ($9.49$). Same as the graph size, graphs within each category are sampled using Gaussian distributions centered on the category median.
 
\subsubsection{Visualization Layout (Layout)}
As we select the node-link diagram as a visual representation of graphs, the layout of the node-link diagram has a great effect on the perception of graphs~\cite{burch-graph-empirical-evaluation, bartolomeo-graph-layout-evalutaion}.
We employ three different graph layout algorithms in different categories based on their popularity and ease of use~\cite{effect-of-graphlayout}: 
\begin{itemize}
    \item \textbf{Force-directed Layout (Fruchterman-Reingold)}: An algorithm which simulates a physical system where nodes act as repelling particles and edges act as attracting springs. This algorithm creates an aesthetically pleasing layout with uniform edge lengths, effectively revealing symmetries and clusters~\cite{forcedirected}.
    \item \textbf{Circular layout}: This layout positions all nodes equidistantly along the circumference of a circle. It provides a structured view that highlights edge density and connectivity patterns across the graph, without the node occlusion~\cite{circular}.
    \item \textbf{Multidimensional scaling layout (UMAP)}: Uniform Manifold Approximation and Projection (UMAP) is a non-linear dimensionality reduction technique. When applied to graphs, it projects the topological structure into a 2D space, effectively preserving both local neighborhood and the global structure~\cite{umap}.
\end{itemize}

\subsubsection{Visual Encodings and Standardization}
To mitigate confounding factors from low-level visual variations, we standardize visual encodings using a systematic scaling strategy. Node sizes and edge widths are proportionally reduced for larger or denser graphs to prevent occlusion, with identical parameters applied uniformly across all layout algorithms to ensure fair comparison.
Regarding color encodings, we prioritize legibility and familiarity to ensure that nodes remain visually distinct even amidst edge clutter.
To achieve this, we adopt the \textit{Tableau10} palette, a color scheme widely used in the information visualization community, assigning the blue hue to nodes and the gray hue to edges.

\paragraph{Summary}
The final dataset included 1,881 node-link diagrams: 1,152 synthetic and 729 real-world graphs. Although some conditions fall short of the target of 32 graphs due to their inherent sparsity and multi-component structures (particularly very large and very dense cases), sufficient diversity in visual comparisons is maintained. Specifically, except for the very large–very dense condition, which yielded no graphs, at least six graphs were collected for all other conditions, allowing for a sufficiently diverse set of comparisons. Dataset can be found online\endnote{Github repository, \href{https://github.com/skwn-j/gsa-dataset}{https://github.com/skwn-j/gsa-dataset}}.

\begin{figure*}[t]
  \centering
  \includegraphics[width=\textwidth, alt={layouts}]{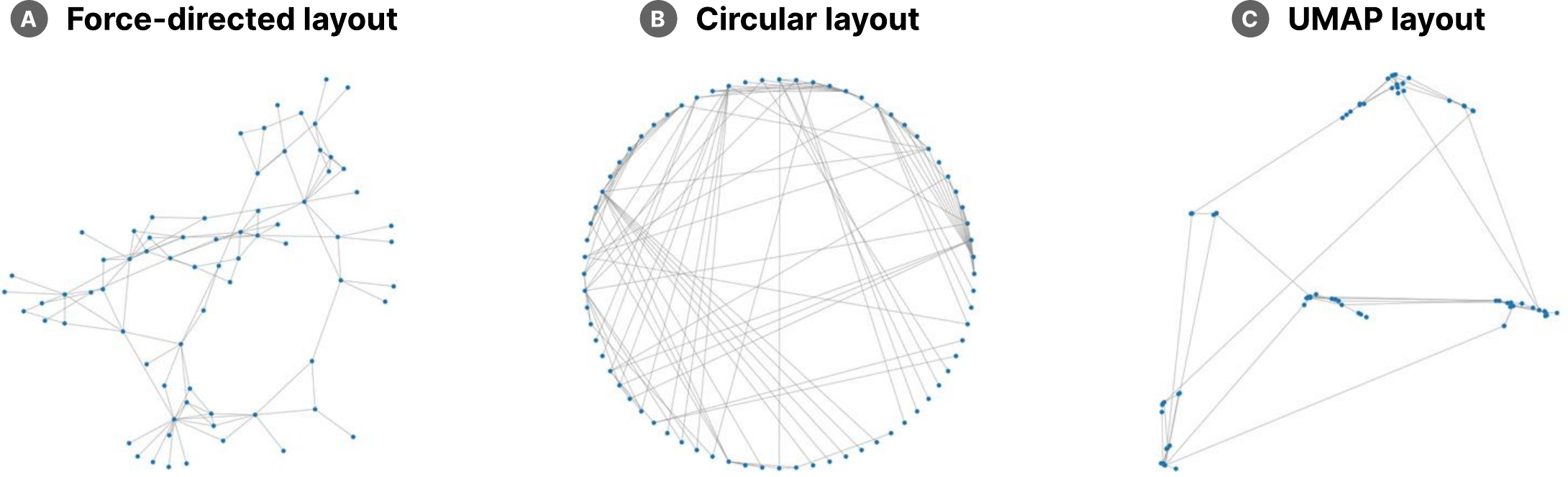}
  \caption{
    Node-link diagrams of a real-world graph drawn with three different graph layout algorithms utilized in this study. A) \textbf{Force-directed layout (Fruchterman-Reingold)} creates an aesthetically pleasing layout with uniform edge lengths by simulating a physical system where nodes act as repelling charged particles and edges act as attracting springs~\cite{forcedirected}. B) \textbf{Circular layout} positions all nodes equidistantly along the circumference of a circle, providing a structured view that highlights edge density and connectivity patterns across the graph~\cite{circular}. C) \textbf{Multidimensional scaling layout (UMAP)} projects the topological structure into a 2D space with the UMAP algorithm, effectively preserving both local neighborhood relationships and the global structural organization~\cite{umap}.
  }
  \label{fig:layout}
\end{figure*}

\section{Experiment 1: Assessing Human Perception} 
\label{sec:ex1}
This section describes Experiment 1, including the collection of human subject data from visual graph comparison tasks and the analysis of the results. The primary objective is to answer the first research question: \textit{Can humans discern similarity and differences between graph visualizations, and which factors significantly affect these decisions?}

To answer this question, we first examine whether a consensus exists in human graph similarity assessments that can be evidence that humans can discern the similarity between graphs. Furthermore, we investigate whether human perception extends beyond simple binary differentiation to capturing the varying degrees of similarity. Finally, we identify the specific visual features underpinning these cognitive processes and evaluate how perceptual performance varies across three graph attributes: graph size, density, and layout.

\subsection{Experiment Design}

\subsubsection{Ground Truth Establishment} 
Ground truth is essential for measuring accuracy and evaluating human ability.
However, defining an objective ground truth for graph similarity is challenging due to the lack of a universal mathematical definition or verified computational proxies~\cite{sedlmairTaxonomyVisual2012}. 
In this circumstance, when algorithmic metrics are insufficient, we use human interpretation as a proxy for ground truth~\cite {scatterplotperception}. We establish ground truth using a visual clustering strategy, assuming that intra-cluster similarity exceeds inter-cluster similarity~\cite{jain2010data}. For synthetic graphs, we rely on generation algorithms to distinguish visual clusters because they yield distinct topological patterns (see \autoref{fig:synthetic}). 
Real-world graphs are manually clustered by three authors, including the first author. 
To avoid bias, the authors perform this task based solely on holistic visual resemblance without any predefined criteria or guidelines. 
The individual clustering results are aggregated, and the final groups are established by majority consensus, ensuring that only graphs consistently perceived as similar are grouped together.

\subsubsection{Task Design} 
The experiment employs a triplet, a query graph, and two target graphs for the relative graph comparison task. When given a triplet, participants are asked to select the target most visually similar to the query, 
without any available interactions such as zooming or panning. This restriction is given to maintain strict experimental control by ensuring consistent viewing conditions across all participants, while also prioritizing holistic perceptual judgments over overly detailed analysis.
Then, participants are asked to justify their choice based on six predefined criteria and rate their confidence on a 5-point Likert scale. 

This design moves beyond traditional binary choices to capture nuanced similarity decisions~\cite{clusetme}, adopting a relative comparison task from a previous study on the similarity perception of timeseries visualizations~\cite{comparing-similarity-timeseries}. Confidence ratings allow quantification of perceived similarity, from indistinct differences ($1$) to clear visual discriminability ($5$).

Regarding visual arrangement, we place the query graph at the center and juxtapose the target graphs side by side.
Under the constraints of this study, this choice is the most appropriate for performing the task among other alternatives: superposition and explicit encoding~\cite{considerations-comparison}. Superposition, one of the design alternatives, is infeasible because our dataset comprises graphs without node correspondence, making alignment computationally prohibitive (NP-hard)~\cite{huang2006maximum}. Likewise, we exclude explicit encoding to maintain the study's scope on baseline node-link diagrams without auxiliary visual encodings.

To minimize experimental noise and avoid the decision paralysis observed in pilot studies where both target graphs originated from clusters distinct from the query, we strictly control the questionnaire generation process to ensure that each triplet contains at least one target graph from the same visual cluster as the query graph, thereby guaranteeing a baseline level of visual similarity. Therefore, the triplets are categorized into two conditions: the \textit{Same-Group condition}, where all three graphs are drawn from the same cluster, and the \textit{Distinct-Group condition}, where exactly one target originates from a different group.
Finally, to ensure randomness and eliminate positional bias, the on-screen placement of the targets is randomized for each trial.

\subsubsection{Aligning Graph Orientation}
Furthermore, the questionnaires are designed to minimize errors arising from the stochastic nature of graph layout algorithms and the inherent characteristics of human visual perception. Layout algorithms, particularly force-directed models, can produce varying outputs for the identical graph structure~\cite{forcedirected}. Even if the same graph data is presented in an identical layout, variations in orientation alone can cause participants to perceive them as distinct graphs~\cite{colinware}.

Previous research preserves the user's mental map by minimizing node displacement between pairs, which is crucial for accurate similarity assessments~\cite{timelighting, mentalmap-preservation}. However, direct node alignment is infeasible due to the UNC. Alternatively, we employ a computer-vision-based Intersection over Union (IoU) approach ~\cite{inkratio, sticklinks}. We align graph centroids and apply dilation to calculate the IoU Area Under the Curve (AUC).
This metric quantifies the pixel-wise similarity between the graphs. To find the alignment that yields the highest similarity, we iteratively rotate the target graphs by $10^{\circ}$ increments to identify the orientation that maximizes the AUC relative to the query graph.

\begin{figure*}[t]
  \centering
  \includegraphics[width=\textwidth, alt={geospatial}]{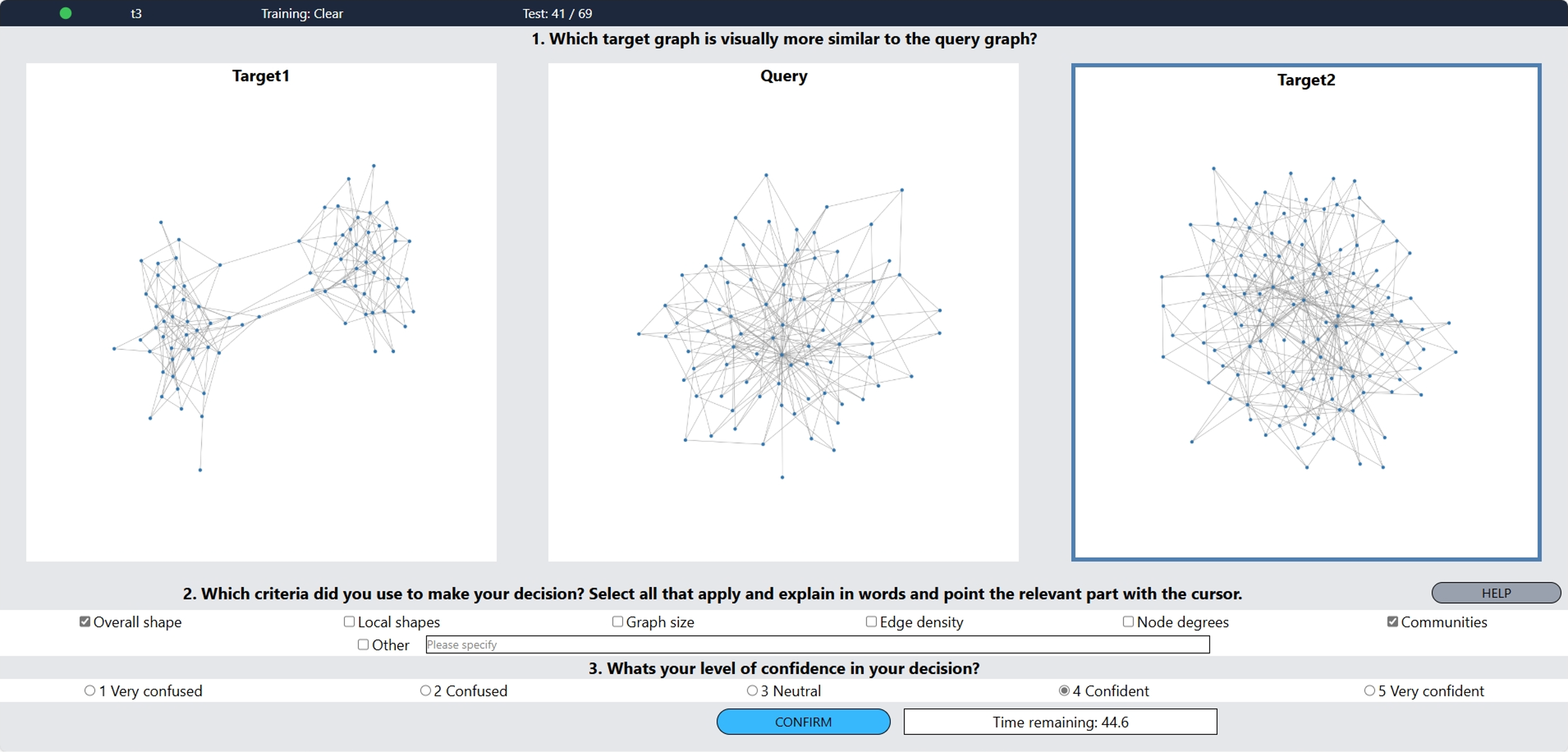}
  \caption{
The system employed in Experiment 1 is designed to collect human graph similarity judgment data. For each question, three node-link diagrams are presented. Participants answer the questions in the following order.1) The user selects the target graph that seems more similar to the central query graph. 2) Next, they choose and explain the decision criteria from the options below. 3) They then indicate their confidence in their choice. 4) The entire process must be completed within one minute. If additional clarification on the criteria is needed, the user can press the Help button to review the explanation.
  }
  \label{fig:system}
\end{figure*}

\subsubsection{Criteria of Human Decisions}
\label{sec:criteria}
Humans employ various visual criteria when assessing graph similarity. To identify the primary criteria used in these decisions, previous studies either encourage open-ended responses with qualitative analysis~\cite{nobre2024reading} or provide predefined criteria to facilitate structured responses~\cite{xiong2019curse}. We adopt the latter method, providing participants with clearly defined criteria derived from existing literature. This structured approach helps participants articulate their reasoning consistently and precisely. The finalized criteria presented to participants are as follows:

\begin{itemize}
\item \textit{Overall Shape}: Global visual arrangement or silhouette of the graph.
\item \textit{Local Shapes}: Specific structural patterns or motifs formed by subsets of nodes and edges.
\item \textit{Graph Size}: Total number of nodes.
\item \textit{Node Degrees}: Distribution and prominence of highly connected (high-degree) nodes.
\item \textit{Edge Density}: The Extent to which edges are densely or sparsely connected among nodes.
\item \textit{Communities}: Identifiable clusters or groups of nodes forming distinct substructures.
\end{itemize}

To develop these criteria, we review previous research investigating visual similarity perception in graphs. Ballweg et al.~\cite{ballweg2018visual} found that participants frequently relied on overall graph shapes, hierarchical structures, and node distribution patterns when judging Directed Acyclic Graphs. Bridgeman and Tamassia~\cite{bridgeman2004user} demonstrated significant influences from global layout consistency, graph size variations, and edge structural changes on perceived graph similarity. Von Landesberger et al.~\cite{investigatin-gs-perception} further supported these findings, noting that local structural variations such as appearance or disappearance of specific connections, edge density, node centrality, and clearly defined community structures significantly influenced similarity decisions.

\subsection{Experiment Process}
This study is approved by the Institutional Review Board of our institution (IRB No. 2501/004-009).
32 adult participants (26 males, 6 females), aged 22 to 35 (mean age 26.6), are recruited. Thirty participants hold university degrees (BSc, MSc, PhD), all reporting intermediate familiarity with graph data and visualizations. Experiments are conducted in-lab or via Zoom using standard desktop or laptop computers with a mouse. Each session lasts approximately one hour, including introductions, training, and tasks. Participants provide informed consent and receive instructions detailing the research procedure. Instructions include an introduction to the tasks participants are required to complete in the three stages, along with explanations and example figures illustrating the criteria to be selected during the second stage.

After the instruction, participants complete three training tasks using additional unused data before the main experiment. 
In the actual experiment, each participant performs two tasks for each source type (real-world and synthetic) across the 36 combinations of independent variables. 
However, due to the exclusion of the three very large and very dense real-world cases, participants complete a total of 69 tasks individually.
Task order is counterbalanced using a Latin square design.

Participants are encouraged to make decisions within 60 seconds per task to avoid overly detailed analysis~\cite{tseng2023measuring}. They are explicitly informed that no guidance is provided on similarity decisions or classification criteria, ensuring the responses genuinely reflect participants' abilities to interpret graphs visually. Instead, participants can revisit the instruction describing the criteria via a help button, and explanations for each criterion are provided immediately upon request (shown in \autoref{fig:system}).

\subsection{Results and Findings}

\subsubsection{Consistency in Human Decision}
\begin{figure}
    \centering
    \includegraphics[width=\linewidth]{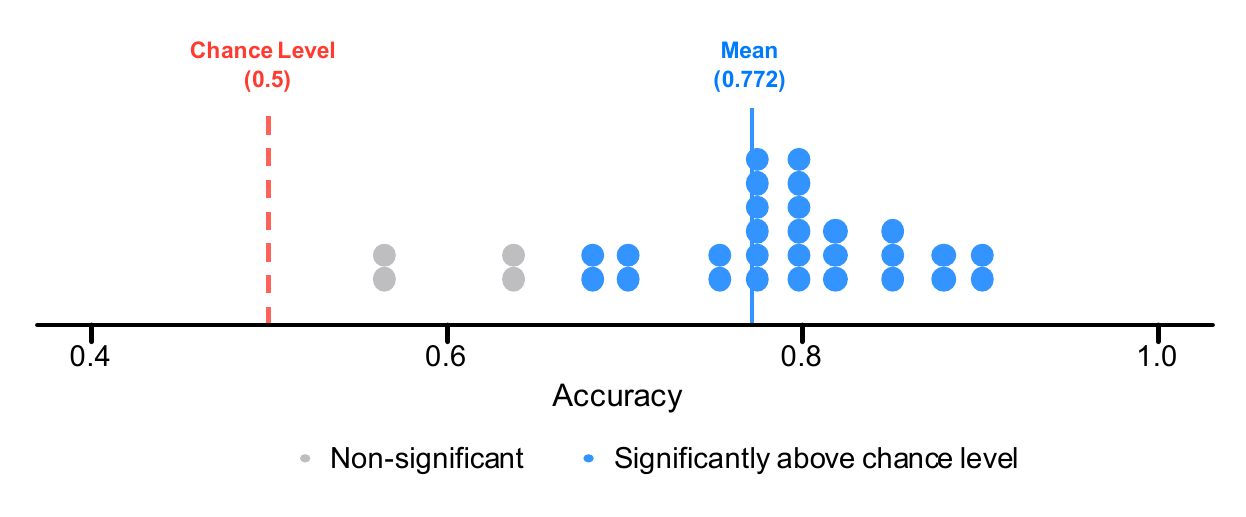}
    \caption{Accuracy distribution of participants in Experiment 1. The red dashed line represents the random chance level (0.5). Each dot represents an individual participant. Based on a one-sample t-test ($H_0 = 0.5$) for each participant, blue dots denote those who exhibited accuracy levels significantly above chance (28 out of 32, $p < .05$), while gray dots represent those whose performance was statistically indistinguishable from chance. These results demonstrate a robust human capacity to visually distinguish graph similarities.}
    \label{fig:accuracy}
\end{figure}

To assess whether humans can perceive visual similarities in graphs consistent with the ground truth, we conduct a one-sample t-test of participants' accuracy against the chance level of 0.5. The analysis reveals that participants achieve a mean accuracy of 77.2\% ($M=0.772, SD=0.086$), which is significantly higher than the random chance level ($0.5, p < .001$). Notably, 28 out of 32 people show accuracy significantly above the chance level, and even the participant with the lowest performance (56.1\%) performs above chance (\autoref{fig:accuracy}). Moreover, the relatively low standard deviation indicates a high level of consistency in visual graph perception across the population. These results strongly support the hypothesis that humans share a consensus on visually distinguishing graph similarities.

To determine which independent variables influence these human decisions, we perform a three-way ANOVA examining the effects of graph size, edge density, and layout on accuracy. The results show that graph size is the only factor exerting a statistically significant main effect on accuracy ($p = .046$). However, post-hoc analysis using Tukey's HSD fails to identify significant differences between any specific pairs of size groups. 
This discrepancy can be attributed to the borderline significance of the ANOVA result and the conservative nature of Tukey's HSD test, which imposes stricter thresholds for pairwise distinctions, indicating that the effect of size is relatively subtle and not driven by distinct disparities across specific groups.
Collectively, these findings imply that, although humans generally struggle to interpret graphs that become excessively large or dense due to factors such as the hairball effect~\cite{survey-emprical-graphvis}, their capacity to perceive graph similarity remains robust, persisting across diverse conditions of size, density, and layout.

\subsubsection{Consistency in Human Confidence}
To further investigate whether participants' visual perception scales with the magnitude of graph differences, we analyze the self-reported confidence scores, ranging from very confused (1) to very confident (5). We hypothesize that participants will report higher confidence in trials in which the target graphs belong to distinct groups (Distinct-Group condition) compared to trials where both targets are from the same group as the query (Same-Group condition). This hypothesis is grounded in the premise that the Distinct-Group condition exhibits more salient visual differences compared to the Same-Group condition, thereby enabling participants to select the target graph from the same group with greater confidence.

To validate this, we verify whether the confidence levels in the Distinct-Group condition are statistically higher than those in the Same-Group condition using a Mann-Whitney U test. The global analysis reveals a statistically significant difference between the two conditions ($U=665,328.0, p < .001$). When calculated on a per-participant basis, participants exhibit higher confidence in the Distinct-Group condition ($M=3.548, SD=0.397$) than in the Same-Group condition ($M=3.221, SD=0.376$), with a mean difference of $0.326$. 

\begin{figure}
    \centering
    \includegraphics[width=\linewidth]{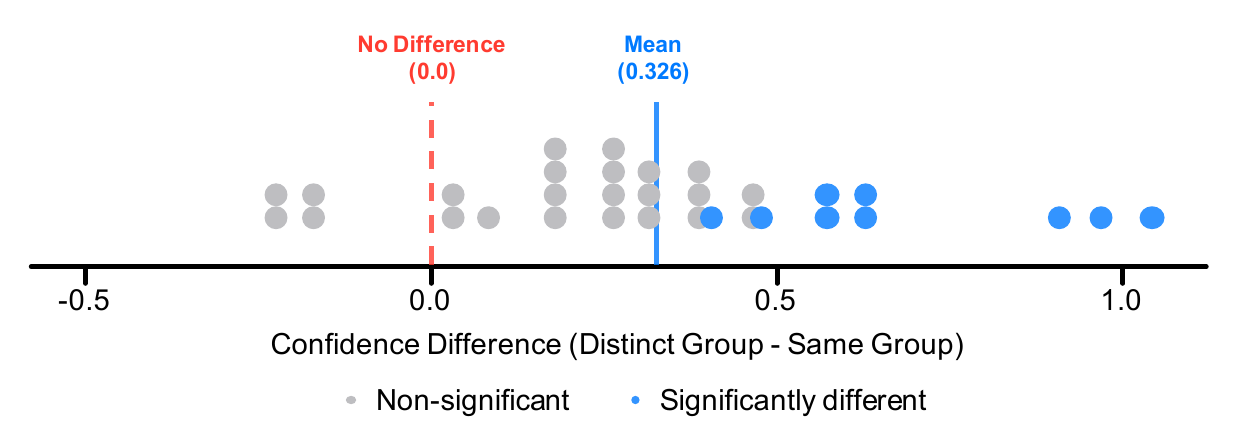}
    \caption{Distribution of confidence differences between Distinct-Group and Same-Group conditions. For each participant, confidence scores (rated on a 1–5 scale) are averaged within each condition, and the mean score of the Same-Group is subtracted from that of the Distinct-Group. The red dashed line marks the baseline of no difference (0.0), and the blue solid line indicates the overall mean difference (0.326). Blue dots represent participants who show a statistically significant difference between two conditions (Mann-Whitney U test, $p < .05$), while gray dots indicate no significant difference. Although the global mean shows a positive trend, the majority of participants (23 out of 32) do not reach statistical significance individually, highlighting substantial heterogeneity in sensitivity to graph differences.}
    \label{fig:confdiff}
\end{figure}

However, a more granular analysis performed at the individual level reveals considerable variability. When independent Mann-Whitney U tests are conducted for each participant, only 9 out of 32 participants ($28.1\%$) demonstrate a statistically significant increase in confidence for the Distinct-Group condition ($p < .05$). The remaining participants do not exhibit a significant difference, and notably, four individuals show inverse trends (\autoref{fig:confdiff}).

These findings suggest that while humans, at a population level, possess the perceptual sensitivity to distinguish not just the similarity but also the magnitude of differences between graphs, this capability is not uniformly distributed. The contrast between the global significance and the individual results highlights a substantial heterogeneity in individual sensitivity to visual graph differences.

\subsubsection{Rationale of Decision}
Participants identify \ovs as the most dominant feature driving their similarity decisions, accounting for the highest proportion of responses ($898$). \eds follows as the second most frequent criterion ($781$), showing a notably higher count compared to other criteria. In contrast, \gsz is the least selected criterion ($227$). These findings suggest that humans prioritize the global silhouette of the graph and readily detect patterns of edge concentration (density), whereas they are less adept at estimating the number of nodes during visual comparison. Total number of decision criteria selected by humans is shown at \autoref{tab:decisioncriteria} with MLLMs' responses.

To investigate the influence of three independent variables on the frequency of each criterion, we perform a three-way ANOVA.
The results reveal distinct interaction patterns: size significantly affects the selection frequency of all criteria except \gsz itself and \cmm. Layout significantly influences all criteria except \lcs and \gsz. Density, in contrast, exhibits a limited scope of influence, significantly affecting only the selection of \eds without impacting the other criteria


\subsection{Takeaways}
The results confirm that humans possess the fundamental capability to visually judge graph similarity, supporting the feasibility of visual comparison tasks. However, while collective human perception reliably reflects the ground-truth difficulty, individual sensitivity to the strength of graph similarity is heterogeneous. Some users are highly sensitive to the magnitude of differences, while others may perform the comparison task accurately but with a flatter confidence profile regardless of task difficulty. Therefore, we recommend designing comparison interfaces that rely on relative judgments rather than requiring absolute quantification of similarity.

Regarding the decision rationale, the overwhelming reliance on \ovs implies that preserving the global topology is paramount in comparative visualization. Systems should prioritize rendering techniques that emphasize the global silhouette. Furthermore, \eds emerges as a critical factor, suggesting that visual patterns created by edge concentrations and resulting occlusions serve as key discriminatory cues.

Among the independent variables, size has the greatest influence on perception, followed by layout and density. Interestingly, accuracy tends to be lower for small graphs. We hypothesize that small graphs lack sufficient node volume to form a distinct global silhouette, making \ovs, the primary human decision criterion, less reliable. In such cases, alternative visual aids may be necessary. Conversely, for medium to very large graphs, the increased number of nodes facilitates the formation of unique global structures, allowing users to effectively distinguish graphs based solely on their \ovs.

\section{Experiment 2: Deriving Human-Measure Relationships}
\label{sec:ex2}

In this experiment, we aim to answer the second research question: \textit{Do traditional computational measures capture the similarities and differences that humans perceive visually?} To address this, we evaluate the alignment between various computational metrics and human decisions. Our objective is to identify measures that not only effectively approximate human perception but also facilitate interpretable comparisons within visual analytics systems. Furthermore, such measures can serve as reliable Visual Quality Measures (VQMs) for graph similarity~\cite{datadriven-evaluation}.

\subsection{Experiment Design}
The selection of computational measures is constrained by the target graph specifications defined in our \hyperref[sec:studydesign]{study design.} Specifically, the measures must support comparisons between undirected, unweighted graphs without node correspondence (UNC). Under these constraints, we select a total of 16 graph similarity measures. A list of the selected measures is provided in~\autoref{table:measures}.

First, we include four graph-attribute-based measures that correspond directly to the quantifiable visual criteria identified in Experiment 1. Among the six criteria used by humans, \ovs and \lcs are excluded due to the difficulty of direct quantification. For the remaining four attributes (Graph Size, Edge Density, Node Degrees, and Communities), we derive similarity scores using two metrics. \textit{Balance}, defined as the ratio of the minimum to the maximum value, is utilized to calculate the similarity of size and density. \textit{Divergence}, calculated as the Jaccard similarity between distributions, is for comparing the distribution of node degrees and sizes of communities detected by the Louvain algorithm.

The remaining 12 measures are selected based on our 
\hyperref[sec:rw-measures]{\textit{literature review}} to ensure a diverse representation of theoretical foundations. Each measure is characterized by distinct properties, including embedding-based~\cite{netsimile, portraitdivergence}, graphlet-based~\cite{gcd, graphletcorrelationdistance}, spectral-based~\cite{netlsd, graphspectra, feather}, graph kernel-based~\cite{graphkernels, grakel}, and graph alignment-based approaches~\cite{alignment-2008}.

To ensure a consistent representation of similarity across all measures, measures that originally denote distances are normalized to a similarity scale in which 0 denotes complete dissimilarity and 1 denotes identity. 
Specifically, distinct transformation rules are applied based on the theoretical range of each metric: for measures strictly bounded within $[0, 1]$, the transformed similarity is calculated as $1-d$. For measures defined on the unbounded range $[0, \infty)$, the transformation $1/(1+d)$ is employed.

\begin{table}[t!]
    \centering
    \caption{List of graph similarity measures employed in Experiment 2. Size and density balances are computed as the ratio of the maximum value to the minimum value. Node degree and community divergences are calculated by computing the Jaccard similarity between the node degrees and the sizes of the communities detected by the Louvain community detection algorithm. The remaining measures are implemented according to the descriptions provided in the references or by using available code. All measures are normalized to yield values between 0 and 1, with higher numbers indicating greater similarity between graphs.}
    \begin{tabular}{m{0.2\linewidth} | m{0.7\linewidth}}
        \toprule
        \textbf{Category} & \textbf{Measure} \\
         \midrule
         \multirow{2}{*}{Attributes} & Size balance, Node degree divergence, \\ 
         & Density balance, Community divergence   \\
         \midrule
         Embedding & Netsimile  \cite{netsimile},  Portrait divergence   \cite{portraitdivergence}  \\
         \midrule
         \multirow{2}{*}{Spectral} & Laplacian spectral  \cite{graphspectra}, Feather  \cite{feather}, \\
           & Ipsen-Mikhailov \cite{ipsen}, NetLSD  \cite{netlsd} \\
         \midrule
         Graphlets & GCD-11   \cite{gcd}, Netdis  \cite{netdis} \\
         \midrule
         {Alignment} & REGAL   \cite{regal},  GRASP     \cite{grasp} \\
         \midrule
         \multirow{2}{*}{Kernels} & Shortest-path kernel        \cite{shortestpath}, \\
           & Weisfeiler-lehman kernel   \cite{weisfeiler}  \\
         \bottomrule
    \end{tabular}
    \label{table:measures}
    \vspace{-0.5cm}
\end{table}

\subsection{Experiment Process} 
Unlike human participants who perform relative comparisons due to the inherent difficulty of absolute quantification, computational measures provide absolute similarity scores for any given pair of graphs. For each triplet consisting of a query graph ($Q$) and two target graphs ($T_A, T_B$), we calculate the computational similarity scores $S(Q, T_A)$ and $S(Q, T_B)$. We then assess whether the choices made by human participants align with the higher similarity score. This process enables us to quantify the concordance rate between algorithms and human visual perception.

\subsection{Results and Findings}
\subsubsection{Agreement with Human Judgment}
To identify the computational measure that best approximates human graph-similarity perception and thus enables interpretable graph comparisons, we analyze the agreement between human judgments and the 16 selected metrics. We employ Cohen's Kappa ($\kappa$) as the primary evaluation metric, as it accounts for the possibility of random agreement, thereby providing a more robust estimate of concordance than accuracy~\cite{cohen-score-valid}. Agreements of all 16 measures are listed at~\autoref{tab:network_measures_colored}.

The analysis reveals that \textbf{Portrait divergence}~\cite{portraitdivergence} demonstrates the best performance. It achieves a $\kappa$ value exceeding 0.4, which is interpreted as a moderate level of agreement according to established benchmarks~\cite{cohen-score-valid}. Several other measures also achieve $\kappa$ values near or above $0.4$, falling within the margin of error of \textbf{Portrait divergence} (\textbf{Ipsen-Mikhailov}~\cite{ipsen}, \textbf{NetDis}~\cite{netdis}, and \textbf{NetSimile}~\cite{netsimile}). In contrast, the remaining measures generally produce values below 0.3, with \textbf{GRASP} recording the lowest ($\kappa = 0.1477$), suggesting a negligible agreement with human perception.

To further validate the reliability of the top-performing measures, we conduct an ANOVA to investigate how the three independent variables, graph size, edge density, and layout, affect their agreement with human decisions. The results indicate that \textbf{Portrait divergence} is not significantly influenced by any of the three variables, demonstrating robust agreement across all experimental conditions. Conversely, other top contenders are affected by specific factors: \textbf{Ipsen-Mikhailov} by layout, \textbf{NetDis} by graph size, and \textbf{NetSimile} by both size and layout.

Additionally, we examine the relative rankings for each participant to account for the inherent heterogeneity of human perception. \textbf{Portrait divergence} achieves the highest stability with an average rank of $4.19$, placing as the best for 9 participants and within the top four for the other 11 participants. Wilcoxon Signed-Rank tests confirm its significant superiority ($p < .05$) over most measures, except for the three other top contenders, verifying its consistency across diverse observers.

In conclusion, \textbf{Portrait divergence} proves to be the most reliable measure, exhibiting superior stability in terms of both the strength of agreement and robustness against variations in graph properties and individual differences.

\begin{table}[t!]
\centering
\caption{Agreement (Cohen's $\kappa$) of 16 measures with human perception in selecting the target graph more similar to the query, and the Spearman's correlation ($\rho$) between human confidence and the absolute difference of measured similarities $|S(Q, T_A) - S(Q, T_B)|$. Measures are ranked in descending order of agreement ($\kappa$). \textbf{Portrait divergence} achieved the highest performance in both metrics (bold). While three other measures (\textbf{Ipsen-Mikhailov}, \textbf{Netdis}, \textbf{Netsimile}) record agreement scores comparable to \textbf{Portrait divergence}, no other measure achieved a correlation score within a similarly competitive range.}
\label{tab:network_measures_colored}
\begin{tabular}{lcc}
\toprule
\textbf{Measure} & \textbf{Agree. ($\kappa$)} & \textbf{Corr. ($\rho$)} \\ \midrule
Portrait divergence      & \cellcolor{blue!60}\textbf{0.4247} & \cellcolor{blue!38}\textbf{0.2685} \\
Ipsen-Mikhailov          & \cellcolor{blue!58}0.4111          & \cellcolor{blue!28}0.1967          \\
Netdis                   & \cellcolor{blue!56}0.4012          & \cellcolor{blue!29}0.2046          \\
Netsimile                & \cellcolor{blue!56}0.3966          & \cellcolor{blue!11}0.0796          \\
Feather                  & \cellcolor{blue!50}0.3576          & \cellcolor{blue!20}0.1421          \\
Shortest-path kernel     & \cellcolor{blue!47}0.3352          & \cellcolor{blue!13}0.0903          \\
Node degree divergence   & \cellcolor{blue!45}0.3196          & \cellcolor{blue!24}0.1704          \\
REGAL                    & \cellcolor{blue!44}0.3151          & \cellcolor{blue!21}0.1499          \\
GCD-11                   & \cellcolor{blue!44}0.3134          & \cellcolor{blue!20}0.1465          \\
Laplacian spectral       & \cellcolor{blue!40}0.2862          & \cellcolor{blue!18}0.1252          \\
Density balance          & \cellcolor{blue!38}0.2727          & \cellcolor{blue!15}0.1093          \\
Community divergence     & \cellcolor{blue!34}0.2401          & \cellcolor{blue!26}0.1861          \\
Size balance             & \cellcolor{blue!30}0.2138          & \cellcolor{blue!13}0.0906          \\
NetLSD                   & \cellcolor{blue!25}0.1749          & \cellcolor{blue!3}0.0182           \\
Weisfeiler-lehman kernel & \cellcolor{blue!24}0.1732          & \cellcolor{blue!7}0.0472           \\
GRASP                    & \cellcolor{blue!21}0.1477          & \cellcolor{blue!14}0.1009          \\
\bottomrule
\end{tabular}
\end{table}

\subsubsection{Correlation with Human Confidence} 
In this analysis, we compute Spearman's correlation ($\rho$) between human confidence ratings and the magnitude of the difference in computational similarity scores. The objective is to verify whether the numerical gap between computational scores reflects the degree of certainty perceived by humans. Correlations of all 16 measures are listed at~\autoref{tab:network_measures_colored}.

A direct comparison presents a methodological challenge due to the disparate nature of the data: human confidence is recorded on a discrete, positive 5-point Likert scale, whereas computational measures yield continuous difference values that can be negative. To reconcile these disparate scales, we apply a systematic transformation process. We first compute the absolute difference between the similarity scores of the two target graphs ($|S(Q, T_A) - S(Q, T_B)|$) to eliminate sign ambiguity. Subsequently, these absolute values are normalized to the $[0, 1]$ range using Min-Max scaling and discretized into five equal-width bins (0.2 intervals) to align directly with the 5-point human confidence scale.

Consistent with the agreement analysis, \textbf{Portrait divergence} exhibits the highest correlation with human confidence, achieving a Spearman coefficient ($\rho$) of $0.2677$. Although this value falls slightly below the conventional threshold of 0.3 typically required to indicate a weak correlation~\cite{schober2018correlation}, it is statistically significantly higher than that of all other competing measures.

To further verify the consistency of this correlation across human participants, we analyze each participant's relative rankings of the measures. \textbf{Portrait divergence} demonstrates superior stability, ranking first for 11 participants and within the top four for 12 other participants, with an average rank of $4.34$. A Wilcoxon Signed-Rank test confirms that it significantly outperforms most other measures, except \textbf{NetDis} and \textbf{Community divergence}.

However, a notable distinction emerges regarding robustness. While the agreement rate of \textbf{Portrait divergence} remains stable across all experimental conditions, its correlation with confidence is sensitive to data characteristics. ANOVA results for each measure indicate that the correlation strength is significantly influenced by graph size and edge density. Nevertheless, despite these dependencies, \textbf{Portrait divergence} remains the most effective computational proxy available for approximating human decision confidence.

\subsection{Takeaways}
The extent to which computational measures align with human visual perception of graph similarity varies significantly across metrics. \textbf{Portrait divergence}, identified as the top performer in both agreement and correlation, incorporates topological characteristics across all structural scales and is applicable to all network types~\cite{portraitdivergence}. 
\textbf{Ipsen-Mikhailov}, which also achieved high agreement, measures global spectral distance for comparing the overall topological structures of two networks~\cite{ipsen}.
This finding suggests that measures reflecting global structures align best with human perception, making them the most effective measures to guide humans' visual graph comparison tasks \textbf{(RQ2)}. This is also consistent with our observation in Experiment 1 that \ovs serves as the overwhelming decision criterion for humans.

However, even the best-performing measures demonstrate only moderate capabilities, falling short of serving as ideal VQMs for graph similarity. While the agreement between human decisions and computational measures is moderate, the correlation between measure differences and user confidence remains weak. This discrepancy may stem from the consistently high confidence observed in \hyperref[sec:ex1]{Experiment 1}, which contrasts with the variable magnitudes of computational differences. In summary, while humans can reliably determine which graph is more similar, they appear less adept at quantifying the exact degree of similarity. Consequently, there exists a clear need for alternative approaches that better bridge this gap.

\section{Experiment 3: Assessing the Potential of MLLMs} 
\label{sec:ex3}
In this final experiment, we address the third research question: \textit{Do MLLMs possess the capability to align with human perception and guide users in graph comparison tasks?}
To answer this, we employ three state-of-the-art general-purpose MLLMs to perform the graph comparison task. We assess their alignment with human perception and investigate their potential advantages over the best-performing computational measure (\textbf{Portrait divergence}) identified in \hyperref[sec:ex2]{Experiment 2}.

\subsection{Experiment Design}
Given that MLLMs exhibit distinct strengths across specialized areas and benchmarks~\cite{vlm-blind}, it is crucial to empirically evaluate their performance specifically within the context of graph similarity assessment. To this end, our evaluation framework mirrors the design of Experiment 1, assessing the visual cognitive capabilities of MLLMs through a relative comparison task. Consistent with the methodology in Experiment 2, we measure agreement and correlation with human judgments to determine whether MLLMs can serve as proxies for human perception and provide effective guidance for users in graph comparison tasks. Furthermore, we analyze the alignment of decision criteria to gain insight into the interpretability and rationale underlying the MLLMs' decisions.

\subsubsection{Model Selection}
The landscape of MLLMs is highly competitive, with rapid advancements from major providers. As of October 2025, Anthropic, OpenAI, and Google are recognized as the dominant market leaders~\cite{ventures2025MidYear2025}. 
Accordingly, we selected the latest flagship general-purpose MLLM from each of these three providers for our evaluation:
\begin{itemize}
    \item \textbf{Google Gemini 2.5 Pro} (hereinafter \textbf{Gemini})
    \endnote{Google, \href{https://ai.google.dev/gemini-api/docs/models\#gemini-2.5-pro}{gemini-2.5-pro (Latest update: June 2025)}}
    \item \textbf{OpenAI GPT-5} (hereinafter \textbf{GPT})\endnote{OpenAI, \href{https://platform.openai.com/docs/models/gpt-5}{gpt-5-2025-08-07}}
    \item \textbf{Anthropic Claude Sonnet 4.5} (hereinafter \textbf{Claude})\endnote{Anthropic, \href{https://www.anthropic.com/claude/sonnet}{claude-sonnet-4-5-20250929}}
\end{itemize}

\subsubsection{Prompting} 
To enable MLLMs to perform visual graph comparison effectively, we develop a structured prompting strategy designed to mimic the human task in Experiment 1. We assign each model the persona of an ``Expert researcher in network visualization and graph drawing'' and instructed it to identify which target graph ($T_1$ or $T_2$) is visually more similar to the Query Graph ($Q$).

To encourage reasoning and improve accuracy, the prompt required the models to follow a four-step Chain-of-Thought (CoT) process:
\begin{enumerate}
    \item \textbf{Internal Evaluation:} Internally compare $Q$, $T_1$, and $T_2$ across six visual features (\ovs, \lcs, \gsz, \ndg, \eds, \cmm) to determine the winner.
    \item \textbf{Rationale Formulation:} Formulate a concise explanation justifying the selection, directly stating the key differences that led to the decision.
    \item \textbf{Confidence Assessment:} Assign a confidence score on a scale ranging from very confused (1) to very confident (5).
    \item \textbf{Decision Criteria Contribution Array:} Provide an array of length six corresponding to the visual features. Each value in the vector should be $\{-1, 0, 1\}$, indicating whether the chosen target was inferior, equal, or superior to the alternative target regarding that specific feature.
\end{enumerate}

To ensure reproducibility and deterministic outputs, the temperature parameter is set to 0 for Gemini and Claude. For GPT, the default setting is used because precise temperature control is not available.

\subsubsection{Input Data and Preprocessing}
Along with the prompt, the MLLMs are provided with the three graph visualization images used in the human trials. To maintain methodological consistency with Experiment 1, we employ the identical protocols for image generation and graph alignment. Furthermore, to ensure consistent input quality and token efficiency, the images undergo standardized preprocessing. Each image is cropped to remove excess whitespace and subsequently padded to achieve a square aspect ratio.

We harmonize image resolution to meet the API specifications for each model. While \textbf{Gemini} and \textbf{Claude} support higher resolutions ($768 \times 768$ and $1092 \times 1092$, respectively), \textbf{GPT} is optimized for $512 \times 512$ pixels. To ensure a fair comparison under identical conditions, all images are resized to a maximum of $512 \times 512$ pixels.

\subsubsection{Output Format}
The models were instructed to generate a structured JSON output containing the following fields:
\begin{itemize}
    \item \texttt{Decision}: The label of the chosen graph ($T1$ or $T2$) which seems more similar to the query graph $Q$.
    \item \texttt{Rationale}: A text string explaining the reason for the decision.
    \item \texttt{Confidence}: An integer score (1--5) implying the confidence of the decision.
    \item \texttt{Criteria}: An array of 6 integers representing contributions of six predefined decision criteria.
\end{itemize}

In addition to the content of the response, we record the inference latency (time elapsed from request to response) to evaluate the practical efficiency of each model.

\subsection{Results and Findings}

\begin{figure}
    \centering
    \includegraphics[width=\linewidth]{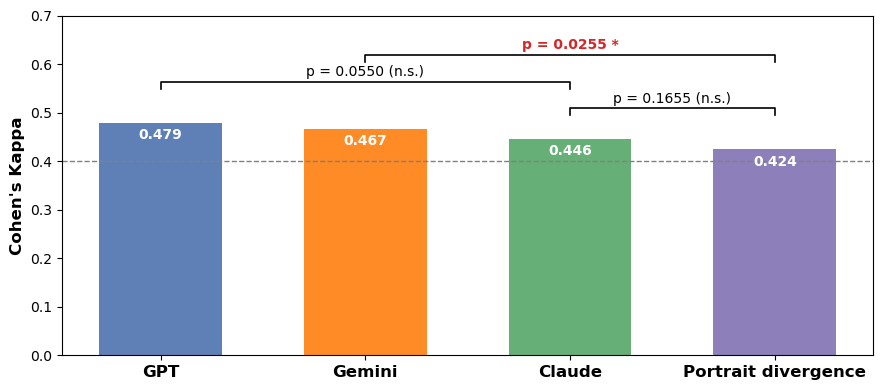}
    \caption{Bar chart comparing the agreement levels (Cohen's Kappa) of three state-of-the-art MLLMs against \textbf{Portrait divergence}. The grey dotted line indicates the moderate level of agreement ($\kappa = 0.4$). Although no statistically significant differences are observed among the MLLMs, \textbf{GPT} achieves the highest nominal performance. Notably, both \textbf{GPT} and \textbf{Gemini} demonstrate significantly higher agreement with human judgments compared to \textbf{Portrait divergence}, indicating that these models effectively surpass the capabilities of traditional computational metrics.}
    \label{fig:llmagreement}
\end{figure}

\subsubsection{Agreement with Human Judgment}
To evaluate the alignment between MLLM decisions and human judgments, we calculate Cohen's Kappa ($\kappa$). Furthermore, to determine whether MLLMs offer a statistically significant improvement over traditional methods, we compare their performance against \textit{Portrait divergence}, which is identified as the top-performing computational measure in Experiment 2. We employ a bootstrap method (resampling $N=2,000$) to test the significance of the difference in Kappa coefficients. Results are shown in~\autoref{fig:llmagreement}.

\textbf{GPT} demonstrates the highest alignment with human judgment ($\kappa \approx 0.479$), showing a statistically significant improvement over \textbf{Portrait divergence} ($\kappa \approx 0.424$, $p=0.0085$). \textbf{Gemini} also exhibits a high level of agreement ($\kappa \approx0.467$), significantly outperforming \textbf{Portrait divergence} ($p=0.030$). In contrast, while \textbf{Claude} achieves a respectable agreement score ($\kappa \approx0.446$), the difference compared to \textbf{Portrait divergence} is not statistically significant ($p=0.168$). Additionally, pairwise comparisons among the models reveal a significant performance gap between \textbf{Claude} and the other two models, whereas the difference between GPT and Gemini was not statistically significant. 

Crucially, this pattern of agreement demonstrates robustness across experimental conditions. To investigate whether the superiority of MLLMs persisted under specific graph attributes, we conduct a stratified bootstrap analysis (resampling $N=2,000$) for each independent variable: graph size, edge density, and layout. Our analysis indicates that while the relative ranking of the methods remained consistent across all variations, the performance gaps between models within individual conditions are not statistically significant. 
Notably, even the difference between the top-performing model (\textbf{GPT}) and the baseline (\textbf{Portrait divergence}) does not reach statistical significance when analyzed within these subdivided categories. Consequently, we interpret the statistically significant superiority observed in the aggregate analysis as the result of marginal performance gains that accumulate consistently across all conditions, rather than being driven by drastic disparities in specific scenarios.

\subsubsection{Correlation with Human Confidence}
\begin{figure}
    \centering
    \includegraphics[width=\linewidth]{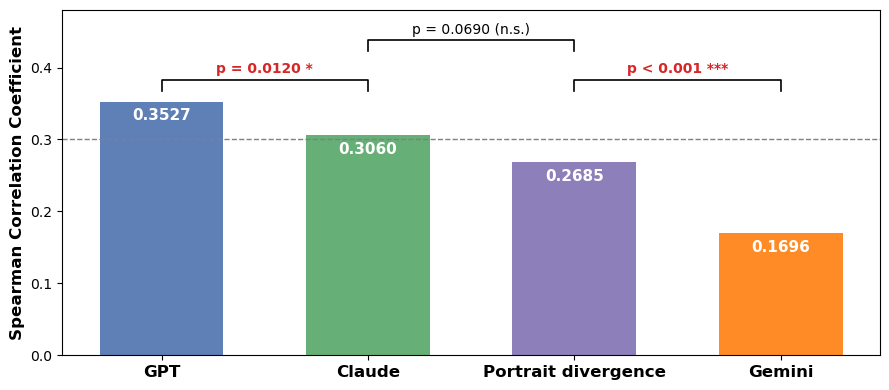}
    \caption{Bar chart comparing the Spearman's correlation ($\rho$) between human confidence and the decision confidence of three MLLMs, benchmarked against \textbf{Portrait divergence}. The grey dotted line indicates a weak correlation level ($\rho = 0.3$). \textbf{GPT} demonstrates a statistically significant improvement in correlation compared to \textbf{Portrait divergence}. \textbf{Claude} shows a comparable correlation with no significant difference observed. In contrast, \textbf{Gemini} exhibits a significantly lower correlation than the computational baseline, indicating a divergence from human uncertainty patterns.}
    \label{fig:llmcorrelation}
\end{figure}

We employ Spearman's correlation coefficient ($\rho$) to assess whether the MLLMs' confidence scores align with the nuances of human certainty. The results, including bootstrap significance tests against \textbf{Portrait divergence}, are presented in~\autoref{fig:llmcorrelation}.

\textbf{GPT} exhibits the strongest correlation with human confidence ($\rho \approx 0.353$), significantly outperforming \textbf{Portrait divergence} ($\rho \approx 0.269$, $p=0.0004$). This suggests that GPT-5 not only identifies the similar graph correctly but also mimics the human distribution of certainty.\textbf{Claude} shows a correlation ($\rho \approx 0.306$) comparable to \textbf{Portrait divergence}, with no statistically significant difference found ($p=0.136$).Conversely, \textbf{Gemini} records a significantly lower correlation ($\rho \approx 0.170$) compared to both \textbf{Portrait divergence} ($p=0.0004$) and GPT-5 ($p < 0.001$).

We hypothesize that \textbf{Gemini}'s lower correlation may be attributed to the temperature setting. To ensure reproducibility, \textbf{Gemini} is queried with a temperature of 0, which tends to force the model into making highly confident, deterministic decisions, thereby reducing the granularity of its confidence scores compared to humans. In contrast, \textbf{GPT}, where precise temperature control is unavailable, likely operates with a default non-zero temperature, resulting in a more natural distribution of confidence scores that better mirrors human uncertainty.

Interestingly, \textbf{Claude} maintains a relatively high correlation ($\rho \approx 0.306$) despite also being queried at a temperature of 0. This contrasts with \textbf{Gemini} suggests that \textbf{Claude} possesses superior internal calibration, allowing it to express nuanced uncertainty even under deterministic decoding strategies. Unlike \textbf{Gemini}, which is prone to overconfidence bias in this setting, \textbf{Claude}'s alignment appears to better preserve the probabilistic granularity akin to humans.

Similar to the agreement analysis, these correlation trends are robust across all conditions of the independent variables. Consequently, the models maintain their respective performance levels regardless of the visual or structural complexity of the target graphs.

\begin{table}[t!]
\centering
    \caption{Criteria selected as decision rationale by humans and MLLMs. While humans tend to select only the most salient features, resulting in a lower total count, MLLMs report a much broader and more frequent range of criteria. The models show strong support for human judgments regarding \ovs, \eds, and \ndg; however, considerable conflicts are observed in \lcs, \cmm, and most notably, \gsz.}
    \label{tab:decisioncriteria}
    \begin{tabular}{llrrr}
    \toprule
    \textbf{Criteria (Human)} & \textbf{Model} & \textbf{Count} & \textbf{TP} & \textbf{FN} \\ 
    \midrule
    \multirow{3}{*}{\shortstack[l]{\ovs \\ (898)}} 
     & GPT    & 1835 & \cellcolor{blue!54}815 & \cellcolor{red!12}83 \\
     & Gemini & 1723 & \cellcolor{blue!53}788 & \cellcolor{red!16}110 \\
     & Claude & 2112 & \cellcolor{blue!59}880 & \cellcolor{red!3}18 \\ 
    \midrule
    \multirow{3}{*}{\shortstack[l]{\lcs \\ (534)}} 
     & GPT    & 1599 & \cellcolor{blue!44}390 & \cellcolor{red!34}144 \\
     & Gemini & 1657 & \cellcolor{blue!48}427 & \cellcolor{red!25}107 \\
     & Claude & 1785 & \cellcolor{blue!49}439 & \cellcolor{red!23}95 \\ 
    \midrule
    \multirow{3}{*}{\shortstack[l]{\gsz \\ (227)}} 
     & GPT    & 921 & \cellcolor{blue!33}124 & \cellcolor{red!58}103 \\
     & Gemini & 1153 & \cellcolor{blue!38}142 & \cellcolor{red!48}85 \\
     & Claude & 1178 & \cellcolor{blue!32}121 & \cellcolor{red!60}106 \\ 
    \midrule
    \multirow{3}{*}{\shortstack[l]{\ndg \\ (541)}} 
     & GPT    & 1950 & \cellcolor{blue!55}492 & \cellcolor{red!12}49 \\
     & Gemini & 1910 & \cellcolor{blue!55}493 & \cellcolor{red!11}48 \\
     & Claude & 1752 & \cellcolor{blue!49}446 & \cellcolor{red!22}95 \\ 
    \midrule
    \multirow{3}{*}{\shortstack[l]{\eds \\ (781)}} 
     & GPT    & 2004 & \cellcolor{blue!56}735 & \cellcolor{red!7}46 \\
     & Gemini & 1991 & \cellcolor{blue!56}725 & \cellcolor{red!9}56 \\
     & Claude & 2007 & \cellcolor{blue!57}741 & \cellcolor{red!7}40 \\ 
    \midrule
    \multirow{3}{*}{\shortstack[l]{\cmm \\ (466)}} 
     & GPT    & 1261 & \cellcolor{blue!49}377 & \cellcolor{red!24}89 \\
     & Gemini & 1123 & \cellcolor{blue!46}359 & \cellcolor{red!29}107 \\
     & Claude & 981  & \cellcolor{blue!43}333 & \cellcolor{red!36}133 \\ 
    \bottomrule
    \end{tabular}
\end{table}

%

\subsubsection{Rationale of Decision}
To evaluate whether MLLMs' reasoning processes align with human perception, we first quantify the frequency with which MLLMs select each decision criterion, and subsequently analyze whether MLLMs identify a specific feature as a decision factor when human participants select it. The frequencies of decision criteria selected by humans and MLLMs, along with their interrelationships, are presented in \autoref{tab:decisioncriteria}.

All MLLMs select the decision criteria significantly more frequently than humans, identifying approximately 2.8 times more criteria on average (Human total: $3,447$ vs. MLLM range: $9,557-9,815$). This disparity likely stems from the differing nature of the reporting task: whereas human participants typically select only the primary or secondary factors influencing their decisions, models typically report every criterion that contributed, even marginally, to their output.

However, distinct variations are observed in the frequency of selection across different criteria. All three models report relatively low frequencies for \gsz ($921-1,178$) and \cmm ($981-1,261$). Conversely, \ovs ($1,723-2,112$), \lcs ($1,599-1,785$), \ndg ($1,752-1,950$), and \eds ($1,991-2,007$) are selected significantly more often. This suggests that, similar to humans, MLLMs regard \ovs and \eds as the most salient factors in visual graph comparison.

Furthermore, the models demonstrate a heightened sensitivity to \lcs and \ndg, features often overlooked by humans. Since these features pertain to specific substructures or local details rather than the overall gestalt, this observation is consistent with findings from related studies suggesting that MLLMs exhibit a more pronounced local focus, whereas human observers emphasize global patterns~\cite{vlm-localfocus}.

When analyzing alignment rates, the generally higher number of criteria selected by MLLMs results in a higher frequency of both mutual agreement (True Positives) and cases in which the model selects a criterion not cited by the human (False Positives). However, conflicts arise when a model fails to identify a criterion that is salient to the human observer (False Negative), potentially causing critical confusion and undermining trust.

Regarding False Negative rates, \ovs ($2.0\%-12.2\%$), \ndg ($8.8\%-17.5\%$), and \eds ($5.1\%-7.1\%$) exhibit relatively low rates. In contrast, \lcs ($17.8\%-27.0\%$) and \cmm ($19.1\%-28.5\%$) show moderate rates, while \gsz demonstrates a distinctively high False Negative rate ($37.4\%-46.7\%$). This discrepancy aligns with the findings from Experiment 2, where the computational measures for \textbf{Size balance} and \textbf{Community divergence} yielded some of the lowest agreement scores. Consequently, these results reinforce the insight that humans are inherently less proficient at visually quantifying nodes or classifying communities compared to detecting topological or density-based features.

\begin{figure}
    \centering
    \includegraphics[width=\linewidth]{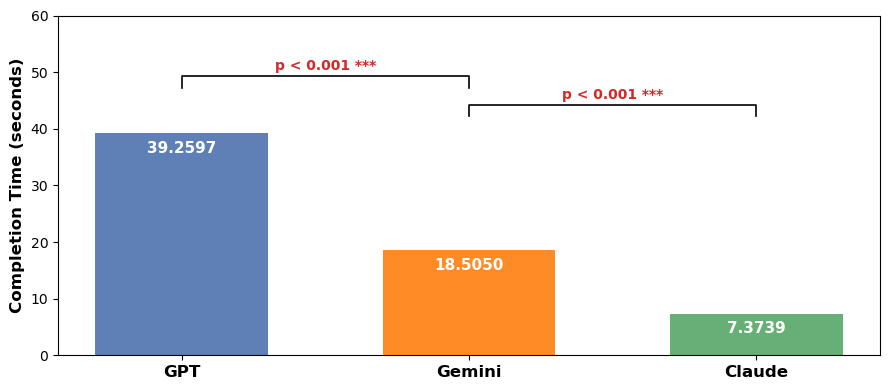}
    \caption{Bar chart depicting the mean of inference latency (completion time) for the three evaluated MLLMs. \textbf{GPT} exhibits the highest computational cost and variance (Mean: 39.26s), highlighting a trade-off between its superior reasoning capability and processing speed. In contrast, \textbf{Claude} demonstrates superior efficiency with the lowest latency (Mean: 7.37s) and a compact distribution. This positions \textbf{Claude} as the optimal choice for time-critical visual analytics scenarios.}
    \label{fig:completiontime}
\end{figure}

\subsubsection{Completion Time}
We analyze the inference latency (completion time) for each model to assess practical feasibility. \textbf{GPT} requires the most computational time (Mean: 39.26s, Max: 142.24s), indicating a trade-off between its high accuracy and latency. \textbf{Gemini} is considerably faster (Mean: 18.50s), completing tasks in less than half the time of \textbf{GPT} on average. \textbf{Claude} demonstrates superior efficiency (Mean: 7.37s), solving tasks significantly faster than both competitors. This positions Claude as a strong candidate for scenarios where real-time processing is critical. Results are shown in \autoref{fig:completiontime}.

\subsection{Takeaways} 
Our experiment results suggest that state-of-the-art MLLMs generally serve as effective proxies for human graph similarity perception, often surpassing the best available computational measures. Grounded on them, we derive the implications for designing visual analytics systems:

For applications that prioritize maximal alignment with human perception, \textbf{GPT} is the optimal choice. It demonstrates statistically significant superiority in both decision agreement and confidence correlation. Although it incurs a higher latency, its ability to provide high-quality rationales alongside accurate judgments makes it a competitive assistant for reducing the cognitive load in complex analysis tasks.

If the primary goal is accurate decision-making (Agreement) rather than calibrating confidence levels, \textbf{Gemini} offers a compelling alternative. It significantly outperforms traditional measures in agreement while being twice as fast as \textbf{GPT}. However, practitioners should be aware of its tendency toward overconfidence (lower correlation).

For time-sensitive or large-scale applications, \textbf{Claude} is the most suitable option. While its accuracy is comparable to that of the best computational measure (\textbf{Portrait divergence}), rather than superior, \textbf{Claude} offers the distinct advantage of providing interpretable textual explanations while maintaining extremely low latency.

Finally, all three MLLMs exhibit consistent reasoning patterns. They provide robust support for human judgments based on \ovs, \eds, and \ndg, whereas they demonstrate relatively high rates of disagreement for \lcs, \cmm, and \gsz. Therefore, when leveraging MLLM reasoning to assist human analysts, we recommend a dual strategy: models can reinforce human decisions regarding the former features, whereas for the latter, they can be utilized to highlight potential discrepancies or aspects that human observers may have overlooked.


\section{Discussion}
In this section, we propose practical guidelines for incorporating MLLMs into visual analytics systems designed for graph comparison with an example scenario. Furthermore, we outline future directions to enhance the generalizability, robustness, and interpretability of graph comparison. Ultimately, we aim to contribute to the efficiency of visual analytics by establishing a more grounded understanding of graph similarity perception.

\subsection{Guidelines for Designing VA Systems for Graph Comparison Tasks with MLLMs}
Based on the empirical findings from our three experiments, we propose the following guidelines for designing visual analytics systems, with a particular focus on applications involving the analysis of temporal evolution and anomaly detection in dynamic graphs.

\subsubsection{Guideline 1. Prioritize Relative Comparison}
Systems should avoid tasks that require users to quantify the exact magnitude of graph changes. Experiment 1 revealed that while humans reliably identify the more similar graph (relative judgment), they struggle to consistently quantify the degree of similarity.
\begin{itemize}
    \item \textbf{Example Scenario:} In a visual analytics system for monitoring network traffic evolution, instead of asking users to ``rate the severity of the change between $t_1$ and $t_2$,'' the interface should present a ranked list of time steps relative to a baseline. This allows users to leverage their relative perception to identify outliers or anomalies effectively.
\end{itemize}

\subsubsection{Guideline 2. Explicitly Encode Quantitative Attributes}
Do not rely solely on node-link diagrams for comparisons involving \gsz or \cmm. Our findings suggest that humans find it difficult to visually identify these attributes in graph visualizations, resulting in the lowest frequency in decision criteria and alignment with computational measures. Fortunately, since these metrics are quantifiable, they should be explicitly represented through auxiliary visualizations rather than through implicit visual encoding.
\begin{itemize}
    \item \textbf{Example Scenario:} When visualizing the temporal evolution of a social network, rather than expecting the analyst to notice a 10\% increase in node or community count, the system should pair the node-link view with a synchronized line chart or bar plot explicitly showing the number of nodes (\gsz) and community counts (\cmm) over time.
\end{itemize}

\subsubsection{Guideline 3. Emphasize Global Structure and Density}
For comparing similar graphs, visualization techniques must prioritize layout stability to preserve the global silhouette~\cite{mentalmap-preservation}. As our findings in Experiment 1 indicate that \ovs and \eds are the overwhelming criteria for human judgment, preserving these features is crucial for emphasizing similarity.
\begin{itemize}
    \item \textbf{Example Scenario:} In a dynamic graph dashboard tracking disease spread, the system should employ mental map preserving layout algorithms. This ensures that the overall shape (\ovs) and density (\eds) of the cluster remain stable across consecutive time steps, allowing analysts to instantly perceive genuine structural changes without additional cognitive load.
\end{itemize}

\subsubsection{Guideline 4. Select Perception-Aligned Metrics for Computational Guidance}
To support human analysts effectively, utilize computational measures that capture global topology, such as \textbf{Portrait divergence}. Employing measures with high alignment improves the interpretability of automated recommendations, as the system's similarity aligns with what the human eye perceives.
\begin{itemize}
    \item \textbf{Example Scenario:} When highlighting significant structural shifts across a sequence of temporal snapshots, the system utilizes \textbf{Portrait divergence} to measure similarity. This ensures that when an analyst drills down to investigate the source of a large discrepancy, the algorithmic difference corresponds to a visually salient feature, allowing the user to verify the change intuitively.
\end{itemize}

\subsubsection{Guideline 5. Strategic Model Selection: Balancing Fidelity and Latency}
Practitioners must balance the trade-off between perceptual fidelity and latency. \textbf{GPT-5} is optimal for in-depth analysis requiring maximal alignment with human intuition, whereas \textbf{Claude Sonnet 4.5} is best suited for real-time monitoring or large-scale screening due to its superior speed and competitive accuracy.
\begin{itemize}
    \item \textbf{Example Scenario:} Use \textbf{Claude} as a real-time filter to flag anomalies, then switch to \textbf{GPT-5} for a comprehensive, post-hoc forensic analysis of the identified events.
\end{itemize}

\subsubsection{Guideline 6. Adaptive Reasoning Strategy: Reinforcement vs. Augmentation}
Systems should adopt a dual strategy for MLLM-generated explanations. Reasoning should reinforce user confidence on global features (\ovs, \eds) where human-model alignment is high, while augmenting analysis by highlighting local details (\gsz, \cmm, \ndg, \lcs) that humans frequently overlook.
\begin{itemize}
    \item \textbf{Example Scenario:} The system validates the user's impression of global shape (Reinforcement) while simultaneously alerting them to local substructure deviations that might be overlooked (Augmentation).
\end{itemize}

\subsection{Expanding the Scope of Graph Comparison}
To improve the generalizability of our findings, future research must broaden the scope regarding graph conditions and similarity measures.
In this study, we focused on the most fundamental graph format---undirected, unweighted, single-component graphs without self-loops---to establish a baseline for comparison. However, even minor topological alterations can significantly impact perceptual outcomes. 
For instance, introducing edge weights necessitates additional visual encodings, such as edge thickness or color. It remains an open question how strongly these salient visual cues might dominate structural perception or alter similarity judgments compared to the unweighted graphs used in this study.

Beyond diversifying graph types, there is substantial room to explore alternative similarity assessment methods. Previous surveys~\cite{masuda2019survey, wills2020survey} indicate that measures leveraging Known Node Correspondence (KNC) tend to offer higher discriminatory power than the Unknown Node Correspondence (UNC) measures used in this work. While KNC was outside our current scope, investigating it would require a fundamental redesign of the experimental framework, such as visibly incorporating node labels to facilitate element-wise comparison. Such an extension would provide deeper insights into how semantic information interacts with topological structure during similarity assessment.

Therefore, future work should aim to replicate these experiments with broader similarity measures and complex graph types. Such expansion is essential to ensure that the proposed guidelines and MLLM proxies remain robust across a wider range of application scenarios.

\subsection{Enhancing the Robustness of Quantitative Analysis}
For our quantitative analysis, we generate a substantial dataset of over 1,000 graph visualizations and collected more than 2,000 human responses. While this volume is sufficient for a preliminary assessment, increasing the sample size is necessary to draw stronger statistical conclusions. Specifically, we encounter challenges in obtaining real-world graphs for certain topological intervals (e.g., very dense, very large graphs), suggesting that data collection methods must extend beyond simple slicing of dynamic graphs to build a truly comprehensive dataset.

Furthermore, the generalizability of our results is constrained by the demographics of our study participants. While the sample size was appropriate for the experimental design, the group exhibited notable homogeneity in terms of prior experience with graph data. This uniformity potentially limits our insight into how similarity judgments vary across different levels of expertise. For instance, it is unclear whether the overwhelming reliance on \ovs observed in this study persists among complete novices, or if experts in specific domains might prioritize different topological features.

Throughout our analyses, the independent variables (size, density, layout) often show limited explanatory power regarding the variance in agreement ($\kappa$) and correlation ($\rho$). This may stem from the inherent subjectivity and heterogeneity of human visual perception. 
However, it is also possible that subtle interaction effects exist but are statistically undetectable due to sample size constraints or participant homogeneity. A larger-scale study with participants from diverse backgrounds would help decouple individual subjectivity from systematic perceptual trends.

\subsection{From Quantitative to Qualitative Understanding}
The primary objective of this study is to quantitatively validate human capabilities in graph similarity assessment and identify computational proxies.
While we successfully identify the criteria selected by participants, our current analysis remains primarily descriptive, focusing on reporting observed patterns rather than explaining the cognitive or perceptual foundations underlying these judgments.

For instance, although participants consistently prioritized \ovs, we haven't theoretically contextualized this finding within established frameworks such as Gestalt psychology or visual attention mechanisms. Consequently, it remains unclear precisely why participants perceive certain global features as salient or how specific geometric properties triggered these decisions.
Similarly, for MLLMs, further investigation is needed to link their performance to internal representations, such as attention maps, to fully understand their visual reasoning processes.

To address these gaps, a rigorous qualitative investigation of textual reasoning from both humans and MLLMs is required. By systematically analyzing the generated explanations, future research can identify the specific visual cues that drive similarity judgments and map them to theoretical concepts. 
Synthesizing this qualitative depth with our quantitative findings will establish a robust theoretical foundation for graph similarity perception in both human and machine agents.


\section{Conclusion}
This study presents a comprehensive empirical investigation into the alignment between human visual perception and machine-driven assessments of graph similarity. By constructing a diverse dataset and collecting extensive human judgment data, we establish that human similarity perception is driven primarily by global visual features such as overall shape and edge density, rather than microscopic topological properties or graph attributes.

Our benchmarking reveals the limitations of traditional computational measures; even the top-performing metric, \textbf{Portrait divergence}, captures human perception only moderately. In contrast, our evaluation of state-of-the-art MLLMs highlights a possibility of better alternatives. Models like \textbf{GPT-5} not only exhibit superior alignment with human judgments but also offer the distinct advantage of explainability through natural language rationales.

While challenges regarding latency and consistency remain, our findings suggest that MLLMs have the potential to become powerful components in the next generation of visual analytics systems.
By serving not only as effective proxies for human perception but also as augmented observers capable of uncovering structural nuances that may elude human eyes, they enable more intuitive, user-centric tools that can both perceive data as analysts do and reveal deeper insights.

\begin{acks}
The authors gratefully acknowledge the support of the BK21 Global Visiting Faculty Fellowship, which facilitated this joint research.
\end{acks}






\begin{funding}
This work was supported by the InnoCORE program of the Ministry of Science and ICT (N10250156), National Research Foundation of Korea (NRF) grant funded by the Korean government (MSIT) (No. 2023R1A2C200520911), the Institute of Information \& Communications Technology Planning \& Evaluation (IITP) grant funded by the Korean government (MSIT) [NO.RS-2021-II211343, Artificial Intelligence Graduate School Program (Seoul National University)], and by the SNU-Global Excellence Research Center establishment project. The ICT at Seoul National University provided research facilities for this study.
\end{funding}

\begin{sm}
Supplemental material for this article is available online.
\end{sm}



\theendnotes

\bibliographystyle{SageV}
\bibliography{refs}

%


\subsection{Copyright}
Copyright \copyright\ \volumeyear\ SAGE Publications Ltd,
1 Oliver's Yard, 55 City Road, London, EC1Y~1SP, UK. All
rights reserved.










\appendix
\newpage
\onecolumn

\lstset{
    basicstyle=\ttfamily\small, 
    breaklines=true,            
    breakatwhitespace=false,    
    frame=single,               
    columns=fullflexible
}

\section{Appendix: LLM Prompt}
\begin{lstlisting}
# Role 
You are an expert researcher in Network Visualization and Graph Drawing. Your task is to evaluate visual similarity between node-link diagrams.

# Instruction
I will provide three images of node-link diagrams:
1. **Query Graph (Q)**: The reference graph.
2. **Target Graph 1 (T1)**: The first candidate for comparison.
3. **Target Graph 2 (T2)**: The second candidate for comparison.

**Your Goal:**
Identify which target graph (T1 or T2) is visually more similar to the Query Graph (Q). Instead of listing all details, provide a concise justification focusing *only* on the decisive factors. You must also **self-evaluate your confidence** in this selection based on how distinguishable the similarity is. Finally, quantify the contribution of each visual feature.

# Visual Features Definitions
Evaluate the graphs based on the following 6 strictly defined features. The order is fixed for the output array.

1. **Overall Structure (Global Topology):** The macro-level shape (e.g., ring, star, cluster).
2. **Substructure (Local Patterns):** Recurring small motifs (e.g., triangles, cliques).
3. **Graph Size (Node Count):** Visual estimation of the number of nodes.
4. **Node Degrees (Hubs vs. Leaves):** Distribution of connections (hubs or uniform).
5. **Edge Density (Clutter):** Visual darkness or hairball-likeness.
6. **Number of Communities (Clusters):** Number of visually distinct groups.

# Output Requirements

Step 1. **Internal Evaluation:** Internally compare Q, T1, and T2 across the 6 features to determine the winner.
Step 2. **Rationale Formulation:** Formulate a concise explanation that justifies why the winner was selected. Directly state the key differences that led to the decision.
Step 3. **Confidence Assessment:** Assign a confidence score (1-5) to your decision based on the following scale:
- `1`: **Very confused** (The difference is negligible; almost a random guess).
- `2`: **Confused** (Hard to distinguish, low certainty).
- `3`: **Neutral** (There are differences, but the decision is borderline).
- `4`: **Confident** (The winner is clearly more similar based on visual evidence).
- `5`: **Very confident** (The winner is obviously identical or extremely similar to Q).
Step 4. **Feature Contribution Array:** Provide an array of length 6: [v1: Overall Structure, v2: Substructure, v3: Graph Size, v4: Node Degrees, v5: Edge Density, v6: Number of Communities].
- `1` (Positive): Winner is *more* similar to Q than the Loser.
- `-1` (Negative): Winner is *less* similar to Q than the Loser.
- `0` (Neutral): Both are equally similar or dissimilar.

# Final Output Format
Please strictly follow the JSON schema provided. The output must be a single JSON object with the following fields:
- selected: ['T1' or 'T2']
- rationale: [Concise explanation text]
- confidence: [Integer 1-5]
- features: [Array of 6 integers as defined above]
\end{lstlisting}
\newpage
\twocolumn

\end{document}